\documentclass[preprint,10pt]{elsarticle}
\usepackage{graphicx}
\usepackage{amsmath}
\usepackage{subfigure}
\usepackage{longtable}
\usepackage{listings} % ���Ӻ���
\newif\ifCLASSOPTIONpagebackref \CLASSOPTIONpagebackreftrue %Reference cross
\newif\ifCLASSOPTIONfinal  \CLASSOPTIONfinaltrue  % final version
\usepackage{amssymb}
\usepackage{amsthm}

\RequirePackage[top=0.5cm,bottom=1cm,left=2.0cm,right=2.0cm,includehead,includefoot]{geometry}
\RequirePackage[rgb,x11names]{xcolor}
\usepackage{indentfirst}

\journal{Communications in Computational Physics}
\begin{document}
\begin{frontmatter}



\title{A moving mesh finite difference method for non-monotone solutions of non-equilibrium equations in porous media}

\author{Hong Zhang$^*$, \quad Paul Andries Zegeling}
\address{Department of Mathematics, Faculty of Science, Utrecht University, Budapestlaan 6, 3584CD Utrecht, The Netherlands }
%\author[lab2]{Hong Zhang$^*$}
%\author[lab2]{Paul Andries Zegeling}
%\author[lab3]{Anika Remorie}
%\address[lab2]{Department of Mathematics, Faculty of Science, Utrecht University, Budapestlaan 6, 3584CD Utrecht, The Netherlands }
%\address[lab3]{address }

\cortext[]{Corresponding author. \newline
E-mail addresses: H.Zhang4@uu.nl (H. Zhang), P.A.Zegeling@uu.nl (P. A. Zegeling)}
\begin{abstract}
An adaptive moving mesh finite difference method is presented to solve two types of equations with dynamic capillary pressure effect in porous media. One is the non-equilibrium Richards Equation and the other is the modified Buckley-Leverett equation. The governing equations are discretized with an adaptive moving mesh finite difference method in the space direction and an implicit-explicit method in the time direction. In order to obtain high quality meshes, an adaptive time-dependent monitor function with directional control is applied to redistribute the mesh grid in every time step, then a diffusive mechanism is used to smooth the monitor function. The behaviors of the central difference flux, the standard local Lax-Friedrich flux and the local Lax-Friedrich flux with reconstruction are investigated by solving a 1D modified Buckley-Leverett equation. With the moving mesh technique, good mesh quality and high numerical accuracy are obtained. A collection of one-dimensional and two-dimensional numerical experiments is presented to demonstrate the accuracy and effectiveness of the proposed method.
% the indicate that the proposed moving mesh FD method could be a suitable to approximate two-phase flows in porous media.
%which can handle the nonlinearities of the governing equations in an efficient way is adopted used to redistribute more grid points near the critical regions, which enables us to obtain more accurate numerical solutions with fewer grid points. 
%For the non-equilibrium Richards equation, theoretical research in \cite{nieber2005dynamic} shows that a propagating water front that is uniform in the lateral direction is conditionally unstable to finite perturbations to the flow field. Our numerical experiments reproduce the similar results: small perturbations start to grow if the wave number of the perturbations is small enough and the parameter $\tau$ is large enough.
\end{abstract}
\end{frontmatter}

%{\it AMS code: \rm 34C55; 35L75; 65M06; 65M20; 76T10.}
%, hysteresis, non linear high order hyperbolic,  finite difference, line method two phase flow,

%% INTRODUCTION %%
\section{Introduction} \label{sec:intro}
For the past several decades, since the observations of saturation overshoot and gravity driven fingers \cite{hill1972wetting,selker1992fingered,glass1989mechanism,dicarlo2004experimental}, there have been a great deal of experimental and theoretical studies on the mechanism and modeling of such phenomena. Stauffer \cite{stauffer1978time}, Hassanizadeh and Gray \cite{hassanizadeh1993thermodynamic}, Kalaydjian et al. \cite{kalaydjian1992dynamic} proposed a dynamic (non-equilibrium) relationship between capillary pressure and saturation to explain the occurrence of non-monotone saturation and capillary pressure when water is injected into initially dry sandy porous media. Eliassi and Glass investigated three additional forms referring to as a hypodiffusive form, a hyperbolic form and a mixed form in \cite{eliassi2001continuum}, they obtained saturation overshoot successfully by using the hypodiffusive form \cite{eliassi2003porous}. Nieber et al. \cite{nieber2003non}, Chapwanya and Stockie \cite{chapwanya2010numerical} investigated the gravity-driven fingers by supplementing the Richards equation with the dynamic capillary pressure-saturation relationship, as well as including hysteretic effects. There results demonstrate that the non-equilibrium Richards equation is capable of reproducing realistic fingers for a wide range of physically relevant parameters. Inspired by fingering instabilites in the flow of thin films, Cueto-Felgueroso and Juanes \cite{cueto2009phase} put forward a phase field model using the idea of including the effect of a macroscopic interface in the mathematical description of unsaturated flow. Their model predictions agreed well with the lab measurements \cite{dicarlo2004experimental}. In the above mentioned references, most of models can be described as extensions to the Richards equation, besides, other approaches characterizing the saturation overshoot have also been investigated. Refs. \cite{hilfer2012nonmonotone, doster2010numerical} studied a generalized theory by introducing percolating and non-percolating fluid phases into a traditional mathematical model. DiCarlo et al. \cite{dicarlo2012fractional} developed a multi-phase, fractional flow approach to describe the physics behind the displacement front that includes the viscosity of the gas. Refs. \cite{van2013non,zegeling2015adaptive,zhang2016mimetic} simulated saturation overshoot by incorporating the dynamic capillary pressure with a traditional fractional flow equation. Their results suggest that the non-equilibrium fractional flow equation has the ability to model saturation overshoot.

Among the proposed theories, two models incorporating the dynamic capillary pressure relationship have attracted considerable interest in recent years. One is the relaxation non-equilibrium Richards equation (RNERE), and the other is the modified Buckley-Leverett equation (MBLE). Results on stability, traveling wave (TW) solutions, global existence, phase plane analysis and uniqueness of weak solutions are given in \cite{egorov2003stability,nieber2005dynamic,van2007new,mikelic2010global,spayd2011buckley,cao2015uniqueness}. Numerical simulations \cite{nieber2003non,chapwanya2010numerical,van2007new,kao2015fast,zhang2016numerical} of the RNERE and the MBLE show that with appropriate parameters, both models will generate non-monotonic distribution of saturation, and the RNERE can become unstable in 2D when the flow profiles are sufficiently non-monotonic \cite{nieber2005dynamic}. 
 
In order to numerically solve these non-equilibrium equations, a variety of numerical methods have been developed in literature. Peszynska and Yi \cite{peszynska2008numerical} proposed a cell-centered finite difference method and a locally conservative Eulerian-Lagrangian method, but they noticed that such methods may cause instabilities in convection-dominated cases and for large dynamic effects. A finite difference method which combined a minmod slope limiter based on the first order upwind and Richtmyer's schemes was used by van Duijn et al. \cite{van2007new}. The solutions obtained by this scheme agreed well with the TW results. Wang and Kao \cite{wang2013central} extended the second and third order central schemes to capture the nonclassical solutions of the MBLE. Kao et al. \cite{kao2015fast} split the MBLE into a high-order linear equation and a nonlinear convective equation, and then integrated the linear equation with a pseudo-spectral method and the nonlinear equation with a Godunov-type central-upwind scheme. The computed solutions demonstrate that the higher-order spatial reconstruction using fifth-order WENO5 scheme gives more accurate numerical solutions. Hong et al. \cite{hong2012generalized} adopted a fourth-order central difference scheme to resolve the spatial resolution and a standard fourth-order Runge-Kutta scheme to march the resulting algebraic system in time, they observed high wave number oscillatory waves under certain parametric conditions. But later work by de Moraes et al. \cite{de2016validity} shows that those oscillatory waves do not satisfy threshold for the existence of non-monotonic wave fronts \cite{van2007new}. Thus they suggested to use schemes with nonlinear numerical stability properties to capture the different shock waves, as well as rarefaction waves.%Pop used . Ref.  \cite{abreu2016computing} chose the Richtmyer's flux in a finite volume conservative form to approximate the advection term at the cell center. 

When capturing solutions of the two-phase flow models numerically, one has to deal with the difficulty related to the steep wave fronts or shocks. Thus, extremely dense meshes are required at the steep fronts or shocks in order to produce physically correct solutions. To overcome this difficulty, several adaptive methods have been developed in the past. Hu and Zegeling \cite{hu2011simulating} used a moving mesh finite element method to discretize the RNERE in the space direction. With the moving mesh technique, high mesh quality and accurate numerical solutions are obtained successfully. Dong et al. \cite{dong2014adaptive} combined a mixed finite element method and a finite volume method to handle the nonlinearities of the governing equations efficiently. By adopting the moving mesh method, they obtained accurate numerical solutions with fewer computational resources. Refs. \cite{zegeling2015adaptive,zhang2016numerical} studied the MBLE with adaptive moving mesh finite difference methods, their results show that to achieve the same accuracy, the adaptive methods need around a factor of 4-10 fewer grid points than the uniform grid case. 

Since the moving mesh methods greatly outperform the uniform mesh methods, the objective of the present work is to study the numerical solutions of the non-equilibrium equations using an adaptive moving mesh finite difference method. This method is based on an MMPDE approach \cite{huang1994moving} which works for a general spatial dimension, but we focus only in 1D and 2D in this paper. In order to distribute the mesh points reasonably, we adopt a time-dependent monitor function with directional control \cite{van2010balanced} and a smoothing technique base on a diffusive mechanism \cite{huang1997analysis}.  

%An outline of the paper is as follows
The other parts of the paper are organized as follows. Section \ref{sec:models} introduces the one-phase RNERE and the two-phase MBLE, together with a review of the TW analysis and the stability results. In Section \ref{sec:adaptive} we will present the moving mesh strategy based on a quasi-Lagrangian approach and discretize the system use a finite difference method in the space direction and an IMEX method in the time direction. In section \ref{sec:num}, several one-dimensional and two-dimensional numerical experiments are carried out to demonstrate the effectiveness of the proposed scheme. Finally, Section \ref{sec:conclusions} ends with conclusions and further comments.
\section{Background} \label{sec:models}
%The description of two-phase flow in prous media continues to pose challenges to modeling, analysis and simulation.
In this section, we derive the mathematical models describing the two-phase flow in a homogeneous porous media. For a more detailed derivation, we refer to \cite{van2013non,hilfer2014saturation}. 

Consider a homogeneous porous medium with a constant porosity $\phi$ and a constant intrinsic permeability $K$. One formulation of the traditional macroscopic theory starts from the fundamental balance laws of continuum mechanics for two phases (the wetting phase and the non-wetting phase) inside the porous medium. Denote the saturation of the wetting phase as $u$, then for a fully saturated porous medium, the saturation of the non-wetting phase is $1 - u$. In a two-dimensional situation, the mass conservation equations for the two phases read
\begin{align}
	\label{eqn:masswetting}
	&\frac{\partial(\phi \rho_w u)}{\partial t} + \nabla \cdot (\rho_w \vec{v}_w) = 0, \\
	& \frac{\partial(\phi \rho_n (1 - u))}{\partial t} + \nabla \cdot  (\rho_n \vec{v}_n) = 0,\label{eqn:massnonwetting}
\end{align}
where $\rho_\alpha$ and $\vec{v}_\alpha, \alpha = n, w$ denote the density and the volumetric velocity of each phase.
Let $z$ be the vertical coordinate taken as positive upward, then Darcy's law reads
\begin{equation}
	\begin{aligned}
		\label{eqn:darcy}
		\vec{v}_\alpha &= -\frac{k_{r, \alpha} K}{\mu_\alpha}(\nabla p_\alpha + \rho_\alpha g \vec{e}_z), \\ 
		& = -\lambda_\alpha (\nabla p_\alpha + \rho_\alpha g \vec{e}_z), \quad \alpha = n, w,
	\end{aligned}
\end{equation}
where $g$ is the gravitational acceleration constant, $\vec{e}_z$ is the unit vector in the $z$ direction, $k_{r \alpha}$, $\mu_\alpha$, $p_\alpha$ and $\lambda_\alpha$ are the relative permeability function, viscosity, pressure and mobility of phase $\alpha$, respectively.
Under non-equilibrium conditions, Stauffer \cite{stauffer1978time}, Hassanizadeh and Gray \cite{hassanizadeh1993thermodynamic}, Kalaydjian \cite{kalaydjian1992dynamic} proposed that the phases pressure difference $p_n - p_w$ can be written as a function of the equilibrium capillary pressure minus the product of the saturation rate of the wetting phase with a dynamic capillary coefficient $\tau$ [Pa s]:
\begin{align}
\label{eqn:dynacapi}
p_n - p_w = P_c(u) - \tau \frac{\partial u}{\partial t},
\end{align}
where $P_c$ modeling the capillary pressure - saturation relationship under an equilibrium condition, is a smooth and decreasing function of saturation $u$, and $\tau$ can be explained as a relaxation time. We refer to \cite{hassanizadeh2002dynamic} for a review of experimental work on dynamic effects in the pressure-saturation relationship.
%where $P_c(u)$ is the static capillary pressure, $\tau$ is the so-called dynamic capillary coefficient and $\frac{\partial u}{\partial t}$ is the dynamic capillary pressure effect.
\subsection{Basic equations}
\subsubsection{The RNERE}
First, we consider a one phase flow model. When the density of the wetting phase (e.g. water) is much larger than that of the non-wetting phase (e.g. air), it is suggested \cite{hilfer2014saturation} to consider the case $\rho_n = 0$ , $p_n = 0$ and $\vec{v}_n = [0, 0]^T$ as a first approximation. Then the non-wetting phase vanishes from the problem and one is left only with the wetting phase. Assuming $\rho_w$ is constant, combining the mass equation (\ref{eqn:masswetting}), Darcy's law (\ref{eqn:darcy}) and the dynamic capillary pressure relationship (\ref{eqn:dynacapi}) gives the RNERE
\begin{equation}
	\label{eqn:rnere2pdes}
	\left\{
		\begin{aligned}
			&\frac{\partial u}{\partial t} - \frac{\partial}{\partial z} (\frac{1}{\phi}\lambda_w  \rho_w g) + \nabla \cdot [\frac{1}{\phi}\lambda_w \nabla  p] = 0, \\
			&p =  P_c(u) -  \tau \frac{\partial u}{\partial t}.
		\end{aligned}
		\right.
	\end{equation}
Substituting the pressure equation into the saturation equation, we obtain 
\begin{align}
\label{eqn:rnere}
&\frac{\partial (\phi u)}{\partial t} - \frac{\partial}{\partial z} (\lambda_w \rho_w g) + \nabla \cdot [\lambda_w \nabla (P_c(u) -  \tau \frac{\partial u}{\partial t})] = 0.
\end{align}
For a simplification of the notation we write (\ref{eqn:rnere}) as
\begin{align}
	\label{eqn:simform}
	\frac{\partial u}{\partial t} + \frac{\partial }{\partial x} F(u) + \frac{\partial}{\partial z} G(u) + \nabla \cdot [D(u) \nabla u] - \tau \nabla \cdot [H(u)  \nabla \frac{\partial u}{\partial t}] = 0,
\end{align}
where
\begin{equation}
	\begin{aligned}
		%\label{eqn:rnerefuns}
		& F(u) = 0, \quad && G(u) = -\frac{1}{\phi} \lambda_w \rho_w g,\\
		& D(u) = \frac{1}{\phi} \lambda_w P'_c(u), \quad && H(u) = \frac{1}{\phi} \lambda_w.
	\end{aligned} \nonumber
\end{equation}
%\begin{align}
%	\label{eqn:rneres}
%	\frac{\partial u}{\partial t}  + \frac{\partial}{\partial z} K(u) + \nabla \cdot [K(u) \nabla (P_c(u) - \tau \frac{\partial u}{\partial t})] = 0,
%\end{align}
%where the constants $\phi$, and $g$ are absorbed in $K(u)$ and $P_c(u)$.
%
%Under equilibrium conditions, the conventional Richards equation for the flow of water in unsaturated porous media may be written in dimensionless form as
%\begin{align}
%	\label{eqn:re}
%	&\frac{\partial (\phi \rho_w u)}{\partial t} + \nabla \cdot (\lambda_w \nabla P_n(u))  +  \frac{\partial}{\partial z} (\frac{k_{r w}K }{\mu_w} \rho_w g) = 0,\\
%\end{align}
%
%It is widely accepted that the traditional Richards equation for two-phase flow in porous media does not support non-monotone traveling wave solutions for the saturation profile. As a consequence various extensions and generalizations have been recently discussed.  Under non-equilibrium condition, the dynamic capillary pressure effect was suggested by \cite{Hassanizadeh and Gray water resour res 1993, 1990 169}. Substituting (\ref{eqn:dynacapi}) into the system of equation (\ref{eqn:re}) results in the relaxation non-equilibrium Richards equation (RNERE)
\subsubsection{The MBLE}
When the two phases (e.g. water and oil) are incompressible, define the total velocity $\vec{v}_T = \vec{v}_n + \vec{v}_w = [{v}_T^x, {v}_T^z]^T$ and the fractional flow rate of the wetting phase $f_w = \frac{\lambda_w}{\lambda_w + \lambda_n}$, then the velocity of the wetting phase can be expressed by
\begin{align}\label{eqn:vwfrac}
	v_w = f [v_T + \lambda_n  (\nabla (p_n - p_w) - (\rho_w - \rho_n) g)].
\end{align}
Substituting (\ref{eqn:vwfrac}) into (\ref{eqn:masswetting}) and incorporating (\ref{eqn:dynacapi}), we can get a two-phase MBLE as (\ref{eqn:simform}), with
\begin{equation}
	\begin{aligned}
		%\label{eqn:funsmbl}
		& F(u) = \frac{1}{\phi}f_w(u) {v}_T^x, \quad  &&G(u) = \frac{1}{\phi}f_w(u) [ {v}_T^z - \lambda_n(u) (\rho_w - \rho_n) g], \\
		& D(u) = \frac{1}{\phi} \lambda_n(u) f_w(u) P'_c(u), \quad  &&H(u) = \frac{1}{\phi} \lambda_n(u) f_w(u).
	\end{aligned} \nonumber
\end{equation}
%\begin{align}
%	\label{eqn:fullmbl}
%	\frac{\partial u}{\partial t} + \nabla \cdot F(u) +\nabla \cdot [ H(u) \nabla (P_c(u) - \tau \frac{\partial u}{\partial t})] = 0,
%\end{align}
%where the flux $F(u)$ and the capillary induced diffusion \cite{cuesta2006non} $H(u)$ are given by
%
%\begin{align}
%	& F(u) = \frac{1}{\phi}f_w(u) [ \vec{v}_T + \lambda_n(u) (\rho_w - \rho_n) g \vec{e}_z],\label{eqn:flux} \\
%	& H(u) = \frac{1}{\phi} \lambda_n(u) f_w(u)\label{eqn:diff}.
%\end{align}
%
%When $\tau$ is taken to be $0$, which means $p_c = P_c(u)$, Eq. (\ref{eqn:fullmbl}) reduces to the Buckley-Leverett equation
%\begin{align}
%	\label{eqn:bl}
%	\frac{\partial u}{\partial t} + \nabla \cdot F(u) = - \nabla \cdot [ H(u) \nabla (P_c(u))].
%\end{align}

%% Traveling wave analysis %%
\subsection{Traveling wave analysis and non-monotonic solutions} \label{sec:tw}
In this section we apply the TW analysis to show the behavior of the wetting front for various values of $\tau$ for the RNERE and the MBLE. The analysis is performed in one-dimension instead of two-dimensions.

\subsubsection{Traveling wave analysis of the RNERE}\label{sec:twmbl}
%%%%%%%%%%%%%%%%%%%Traveling wave analysis%%%%%%%%%%%%%%%%%
In the $z$-direction, the RNERE reads
\begin{align}
	\label{eqn:rnere1d}
	\frac{\partial u}{\partial t} + \frac{\partial G(u)}{\partial z}  + \frac{\partial}{\partial z} [D(u) \frac{\partial u}{\partial z}] - \tau \frac{\partial}{\partial  z}[H(u) \frac{\partial^2 u}{\partial z \partial t}] = 0.
\end{align}
By introducing the TW coordinate $\eta = z - s t$ and substituting $u(\eta)$ into (\ref{eqn:rnere}) we obtain a third order ordinary differential equation (ODE)
\begin{equation}
	%\label{eqn:twrnere}
	\left\{
		\begin{aligned}
			& - s u' + [G(u)]' + [D(u) u']' + s \tau [H(u) u'']' = 0,	\\
			& u(-\infty) = u_+, \quad u(\infty) = u_-, \quad u_+ > u_- \in [0, 1],
		\end{aligned}
		\right. \nonumber
	\end{equation}
	where prime denotes differentiation with respect to $\eta$. Integrating this equation over $(\eta, \infty)$ and assuming
\begin{align}
	\label{eqn:rnereorder2bc}
	[D(u)u' + s \tau H(u) u'')] (\pm\infty) = 0 ,
\end{align}
yields the second-order ODE:
\begin{equation}
	\label{eqn:rnereorder2}
	\left\{
		\begin{aligned}
			& - s(u - u_-) + [G(u) - G(u_-)] + D(u) u' + s \tau H(u) u'' = 0, \\
			& u(-\infty) = u_+, \quad u(\infty) = u_-,
		\end{aligned}
		\right.
\end{equation}
with $s$ determined by the Rankine-Hugoniot condition
\begin{align}
	%\label{eqn:rnererhc}
	s = \frac{G(u_+) - G(u_-)}{u_+ - u_-}. \nonumber
\end{align}
Rewrite (\ref{eqn:rnereorder2bc}) as a first order system of ODEs
\begin{equation}
	\label{eqn:rnereorder1}
	\left\{
		\begin{aligned}
			&u' = v, \\
			&v' = \frac{1}{s \tau H(u)} \big[ s(u - u_-) - [G(u) - G(u_-)] - D(u) v\big].
		\end{aligned}
	\right.
\end{equation}
This system has two equilibria:
\begin{align}
	(u, v) = (u_+, 0), \quad (u, v) = (u_-, 0). \nonumber
\end{align}
The Jacobian of (\ref{eqn:rnereorder1}) reads
\begin{equation}
	A = \left[ \begin{array}{cc}
			0 & 1\\
			\frac{s - G'(u) }{s \tau H(u) } & -\frac{D(u)}{s \tau  H(u) }
	\end{array} \right], \nonumber
\end{equation}
and has eigenvalues
\begin{align}
	\label{eqn:eigenvalue}
	\lambda_{\pm} = \frac{1}{2 s \tau H(u) } [-D(u) \pm \sqrt{(D(u))^2 - 4 s \tau H(u)(G'(u) -s )} ].
\end{align}

\begin{figure}[]
	\begin{center}
		{\includegraphics[width=2.5in] {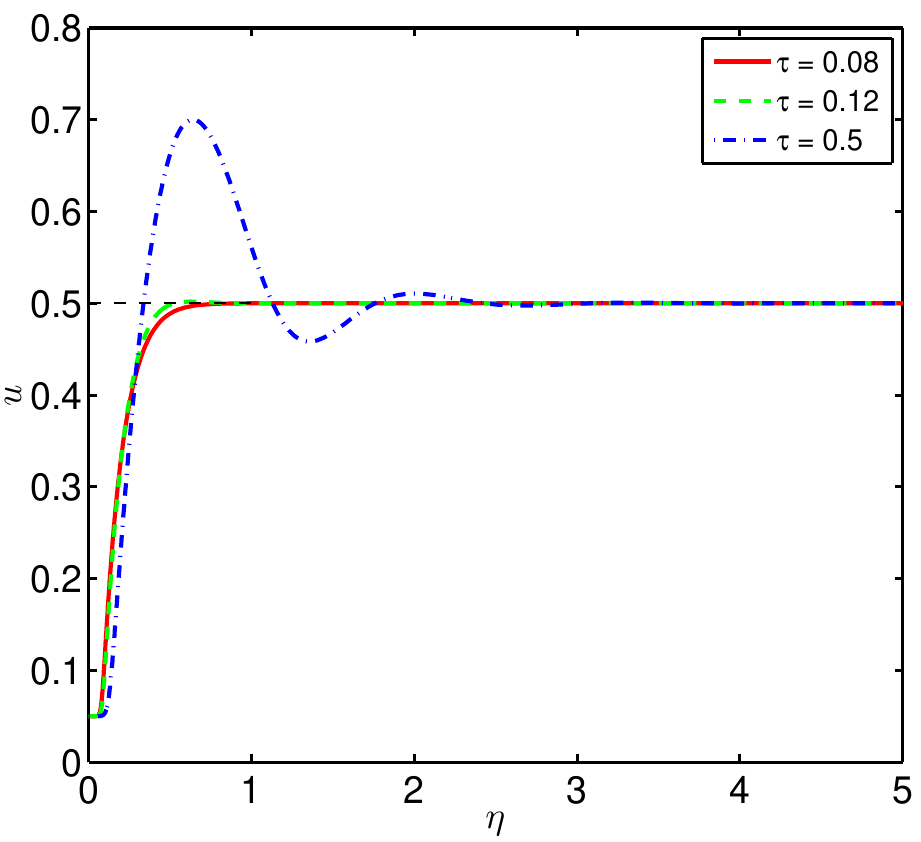}}
		\quad \quad
		{\includegraphics[width=2.5in] {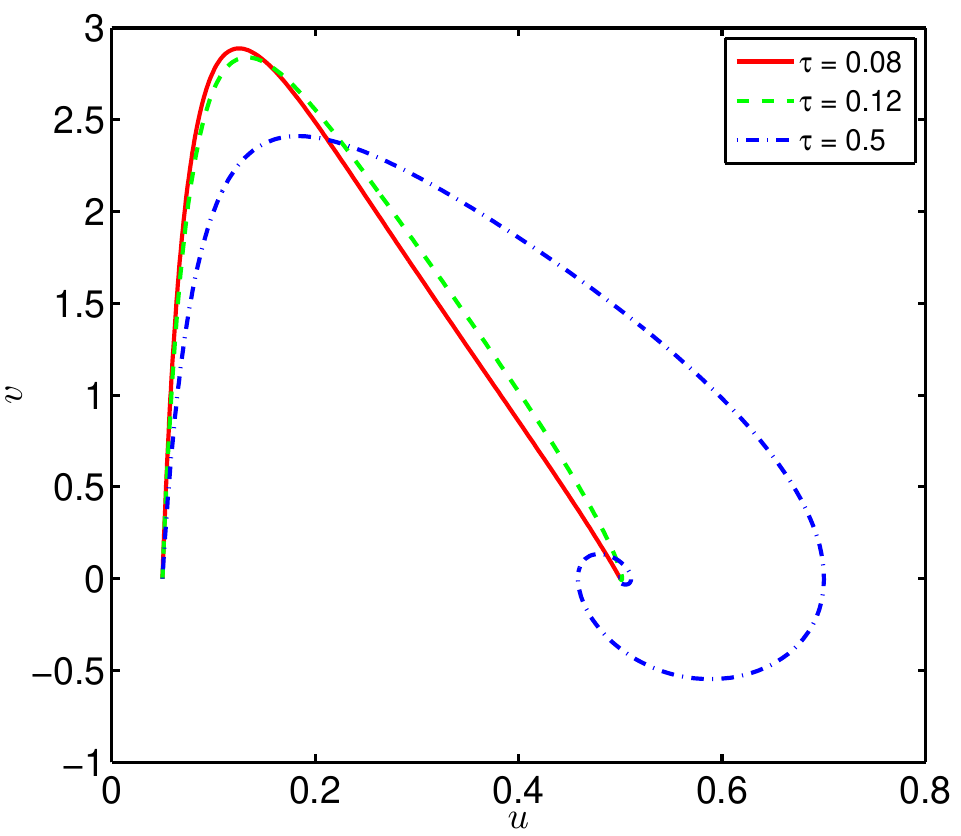}}
		\caption{TW solutions (left) and phase plane plots (right) obtained with $\tau = 0.08$, $\tau = 0.12$, $\tau = 0.5$. \label{fig:niebertwpp}}
	\end{center}
\end{figure}

For the RNERE (\ref{eqn:rnere1d}), TW solutions are possible whenever $u_+ > u_-$. From (\ref{eqn:eigenvalue}) we can get the classification of the two equilibria. The equilibrium $(u_-, 0)$ is a saddle and the equilibrium $(u_+, 0)$ is either an unstable node or a spiral since $G'(u_+) > s$, where the critical value of the dynamic coefficient is defined as
\begin{align}
	\label{eqn:taus}
	\tau_s = \frac{D(u_+)^2 }{4 s H(u) (G'(u_+) - s)}.
\end{align}
When $\tau > \tau_s$, the equilibrium $(u_+, 0)$ is a spiral, which means the saddle point $(u_-, 0)$ is connected to the spiral point $(u_+, 0)$. Fig. \ref{fig:niebertwpp} depicts this situation in terms of the TW profiles (left)  and phase plane plots (right) with the following choice of functions and parameters:
\begin{equation}
	\begin{aligned} \label{eqn:rnerefuns}
		&G(u) = - u^{\alpha}, \quad D(u) = - u^{\alpha - \beta - 1}, \quad H(u) = u^\alpha, \quad \beta = 0.25,\quad \alpha = 3, \\
		&u_+ = 0.5, \quad u_- = 0.05.
	\end{aligned}
\end{equation}
%\begin{equation}
%	\begin{aligned} \label{eqn:vangenuchten}
%		P(u) = - (u^{-1/m} - 1)^{1/n}, \quad K(u) = u^{1/2} \big( 1 - (1 - u^{1/m})^m \big)^2, \quad m = 1 - 1/n, \quad n = 10, \\
%		u_+ = 0.5, \quad u_- = 0.03.
%	\end{aligned}
%\end{equation}
For this choice, (\ref{eqn:taus}) with $u_+$ gives $\tau_s = 0.0843$. If $\tau < \tau_s$, the TW solution varies monotonically (red solid line). With the increment of $\tau> \tau_s$, the TW profile becomes more and more non-monotonic (green dashed and blue dash dotted lines).
\subsubsection{The stability of RNERE} \label{sec:stability}

%following from A Egorov.Stability analysis of traveling wave solution for gravity-driven flow Paragraph 1.
The stability of the RNERE has been discussed in \cite{egorov2002stability,egorov2003stability,nieber2005dynamic}. In a 2D situation, Ref. \cite{egorov2003stability} pointed out that the wetting front for the RNERE is conditionally stable, i.e. stable for high frequency perturbations and unstable otherwise. In this section, we give a summary of the stability results presented in \cite{nieber2005dynamic}.

The stability analysis is based on imposing a small perturbation to the basic TW solution of (\ref{eqn:rnere2pdes}). If the perturbation grows then the flow is unstable. In a 3D domain, the perturbed TW solutions are described as the sum of the basic solutions and the perturbations:
\begin{align}\label{eqn:pertsatuandpres}
	u(x, y, \xi, t) = u_0(\xi) + \epsilon_0 e^{\mathrm{i} \omega_x x + \mathrm{i} \omega_y y + k t} {u}_1(\xi) + \mathcal{O}(\epsilon_0^2), \\
	p(x, y, \xi, t) = p_0(\xi) + \epsilon_0 e^{\mathrm{i} \omega_x x + \mathrm{i} \omega_y y + k t} {p}_1(\xi) + \mathcal{O}(\epsilon_0^2),
\end{align}
where $u_0(\xi)$ and $p_0(\xi)$ are the basic traveling solutions of (\ref{eqn:rnere2pdes}), $\epsilon_0$ controls the magnitude of the perturbation, $\mathrm{i} = \sqrt{-1}$, $\omega = \omega_x^2 + \omega_y^2$ is the wave number of the perturbation with $\omega_x$ and $\omega_y$ being the wave numbers in the $x-$ and $y-$directions respectively. The functions $u_1(\xi)$ and $p_1(\xi)$ describe the variations of solutions and vanish at $\xi = \pm \infty$. The growth factor is denoted by $k$:  if $k$ is positive then the perturbation grows, otherwise it diminishes.

By substituting (\ref{eqn:pertsatuandpres}) into  (\ref{eqn:rnere2pdes}) and dropping the terms of order $\epsilon_0^2$ and higher, the resulting perturbation equations are obtained for ${u}_1$ and ${p}_1$:
\begin{equation}
	\label{eqn:pert}
	\begin{aligned} 
		&\frac{\mathrm{d} A}{\mathrm{d} \xi} + \omega^2 K(u_0) p_1 = -k u_1, \\
		&v \tau_0 \frac{\mathrm{d}u_1}{\mathrm{d} \xi} + (P'(S_0) + v \frac{\partial \tau(u_0,p_0)}{\partial u} \frac{\mathrm{d}u_0}{\mathrm{d} \xi}) u_1 + (v \frac{\tau(u_0, p_0)}{\partial p} \frac{\mathrm{d} u_0}{\mathrm{d} \xi} - 1) p_1 = -k \tau_0 u_1,
	\end{aligned}
\end{equation}
where $A$ is the flux perturbation given by
\begin{align}
	%\label{eqn:pertflux}
	A = -K(u_0) \frac{\mathrm{d} p_1}{\mathrm{d} \xi} - K'(u_0) \big(1 + \frac{\mathrm{d} p_0}{\mathrm{d} \xi} u_1\big) + s u_1,\nonumber
\end{align}
and $s$ is the velocity of the wetting front
\begin{align}
	%\label{eqn:velocity}
	s = \frac{K(u_+) - K(u_+)}{u_+ - u_-}.\nonumber
\end{align}
Nieber et al. \cite{nieber2005dynamic} numerically solved the spectral problem (\ref{eqn:pert}) for various values of $\tau$ and $\omega$. From Fig. 7 in \cite{nieber2005dynamic}, it is observed that when $\tau$ is small enough, the growth factor is negative for all wave numbers and therefore the saturation profile is stable. With increasing $\tau$, the growth factor increases from negative to positive for wave numbers that are not too large. These results on the conditional stability of the RNERE show that the solution can be unstable if the parameters fall within a specified range.
%\begin{figure}[!htbp]
%\begin{center}
%	{\includegraphics[width=2.5in] {./niebergrowthfactor}}
%	\caption{The maximum growth factor $k_0$ as a function of wave number $\omega$. The curve labels 1 to 6 represent different values of $\tau_0$, increasing from $0$ to $1.0$. This plot was   . \label{fig:nieberpert}}
%\end{center}
%\end{figure}

In Section \ref{sec:growthfactor}, we will investigate the stability of the RNERE by numerically solving (\ref{eqn:rnere}) with perturbations and see whether we can observe a comparable behavior for the conditions under which perturbations can grow.
% [K Karlsen.1998.The corrected operator splitting approach applied to a nonlinear advection diffusion problem]
%has the usual s-shaped form, which we mimic using the analytic expression, the capillary diffusion coefficient $D(u)$ is generally a nonlinear (bell-shaped) function of $u$, which we recreate using the simple expression 
\subsubsection{Computation of the growth factor of RNERE} \label{sec:growthfactor}
In numerical simulations, for the purpose of examining the unstable behavior of the RNERE, we add a perturbation to the saturation field at $t = \frac{1}{2}T_{end}$ as
\begin{align}
	%\label{eqn:pert}
	u = u_0 + \epsilon_0 \cos(\mathrm{i} \omega x) \frac{\partial u_0}{\partial z},\nonumber
\end{align}
where $u_0$ is the unperturbed solution computed at $\frac{1}{2}T_{end}$, $\omega$ is the wave number in the $x$-direction. The perturbation to the saturation is the product of a perturbation in the $z$-direction and a cosine shape perturbation in the $x$-direction. 

\begin{figure}[!htbp]
\begin{center}
{\includegraphics[width=2.5in] {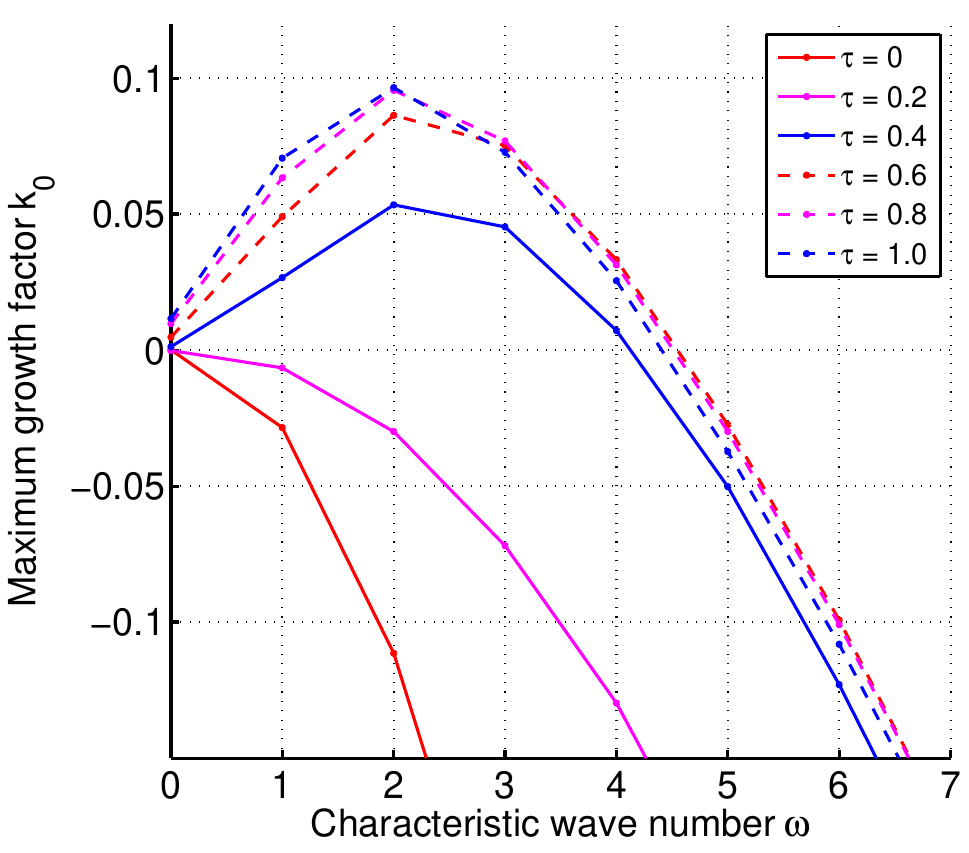}}
{\includegraphics[width=2.5in] {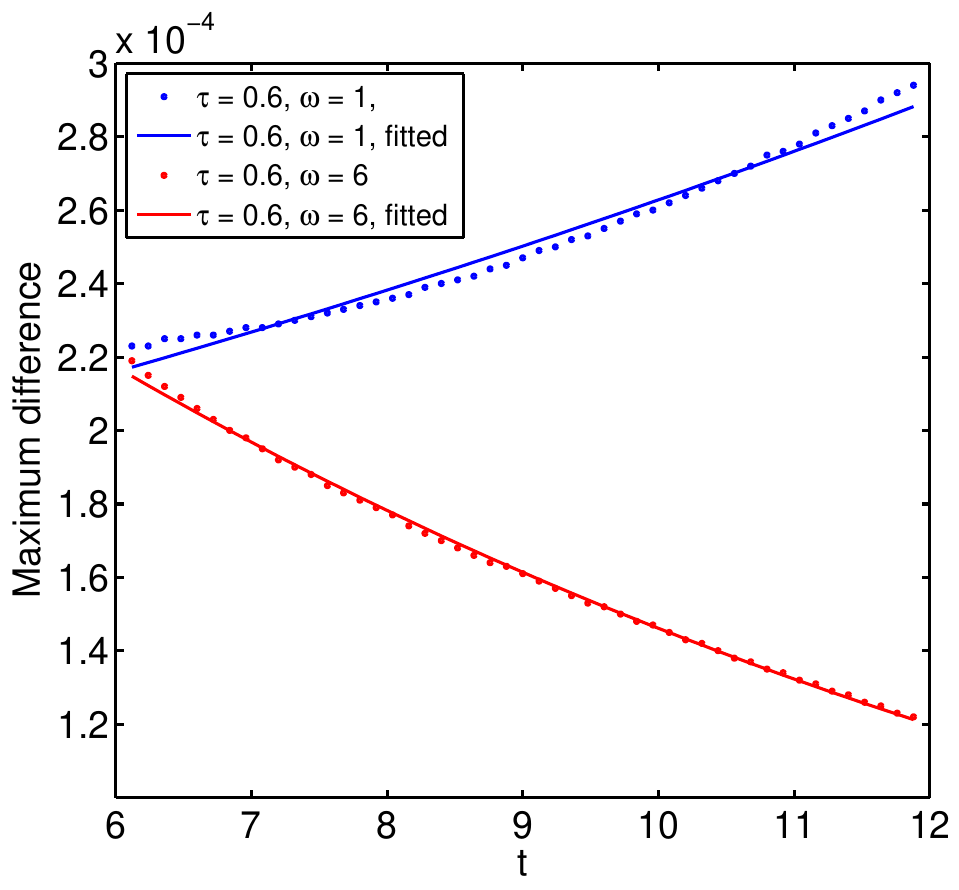}}
\caption{Growth factor vs. characteristic wave number for the RNERE for various values of $\tau$ (left); maximum difference between the perturbed and unperturbed saturation profiles and corresponding fitting curves as functions of time (blue: a growing perturbation with $\tau = 0.6$, $\omega = 1$; red: a declining perturbation with $\tau = 0.6$, $\omega = 6$) (right). \label{fig:growthfactor}}
\end{center}
\end{figure}

In practice, for simplicity only integer values of $\omega$ are considered and the RNERE is solved in one period, which means for wave number $\omega_0$, we set $\omega = 1$ and solve the problem in the physical domain $[0, 2 \pi/\omega_0] \times  [0, 4]$. The first order derivative $\frac{\partial u_0}{\partial z}$ is approximated using the central difference scheme.
%at 50 equally distributed moment during the simulation. 
By solving the RNERE with the perturbed saturation we can determine whether the amplitude of the perturbation increases or decreases in time. The RNERE (\ref{eqn:rnere}) with functions and parameters (\ref{eqn:rnerefuns}) is solved from $t = 0$ to $T_{end} =  12$, first with the unperturbed saturation and then with the perturbed saturation. By subtracting the unperturbed saturation from the perturbed saturation we can get the maximum growth difference. The stability analysis in Section \ref{sec:stability} shows that the evolution of the perturbation has an exponential change, thus we use an exponential least squares fit of the data points to determine the growth factor. Suppose the exponential function that fits the data points is of the following form:
\begin{align}
	y = A e^{t}, \nonumber
\end{align}
where $y$, $A$ and $k$ represent the maximum growth difference, the $y$-intercept and the growth factor, respectively.
Denoting the data points as $(t_i, y_i),~i = 1, 2, \cdots, n$, using the exponential least square fit, we can get $A$ and $k$ as follows:
\begin{align}
	%\label{eqn:expfit}
	&A = \exp( \frac{\sum_{i=1}^n \ln(y_i) \sum_{i=1}^n t_i^2 - \sum_{i=1}^n t_i \sum_{i=1}^n [t_i \ln(y_i)]}{n \sum_{i=1}^n t_i^2 - [\sum_{i=1}^n t_i]^2}), \nonumber\\
	&k = \frac{ n \sum_{i=1}^n [t_i \ln(y_i)] - \sum_{i=1}^n t_i \sum_{i=1}^n \ln(y_i)} {n \sum_{i=1}^n t_i^2 - [\sum_{i=1}^n t_i]^2}. \nonumber
\end{align}
In Fig. \ref{fig:growthfactor} (left) we plot the computed relationship between the maximum growth factor $k_0$ and the wave number $\omega$ for different values of $\tau$. It clearly shows similar behavior as the theoretically plot in Fig. 7 in \cite{nieber2005dynamic}. When $\tau$ is small, the growth factor is negative for all wave numbers. With the increase of $\tau$, the growth factor becomes positive for small wave numbers and the maximum growth factor increases as $\tau$ increases. In Fig. \ref{fig:growthfactor} (right) we plot the data points of the maximum growth difference together with the exponential fitting curves. We can see that the exponential function fits the maximum difference quite well.

%\noindent\textbf{Example 3.} Stability behavior of the RNERE. From examples 1 and 2 we can find that the moving mesh method results in more accurate solutions than the uniform case. Thus we investigate the stability of the RNERE with the moving mesh method. In \cite{egorov2002stability} the growth factor $k$ vs. characteristic wave number for the RNERE are presented. In this example we compare the moving mesh results with the linear stability analysis results.

%\textbf{Conclusion:} From the numerical results, we can see that the RNERE clearly satisfies the stability properties derived theoretically by \cite{A Egorove and J Nieber}
\subsection{Traveling wave analysis of the MBLE} \label{sec:twmble}
The features of the MBLE are richer than the RNERE. In the two-phase situation, the flux function $G(u) = \frac{1}{\phi}f_w(u) [ {v}_T^z + \lambda_n(u) (\rho_w - \rho_n) g]$  is usually a convex-concave function which introduces an additional difficulty to the TW analysis. This case has been extensively investigated in Refs. \cite{van2007new, spayd2011buckley, van2013travelling}, where for a fixed value of $u_-$, the dependency between $\tau$ and the value $u_+$ is analyzed.  For the MBLE the existence of the TW depends on $\tau$. Here we consider $0 < u_- < u_+ < 1$, and let $u_I$ be the unique inflection point of the flux function $G(u)$, we summarize the results as obtained by \cite{van2007new}. For the details of the TW analysis, we refer to \cite{van2007new}.

Similar to the TW analysis of the RNERE, the 1D MBLE in the $z$-direction also has the form (\ref{eqn:rnere1d}) and can be transformed to the ODE (\ref{eqn:rnereorder2}). Consider the following options of $G(u)$, $D(u)$ and $H(u)$:
\begin{align}
	\label{eqn:mblbifu}
	G(u) = \frac{u^2}{u^2 + M (1 - u)^2}, \quad D(u) = -\epsilon, \quad H(u) = \epsilon,
\end{align}
then the results obtained by Ref. \cite{van2007new} can be summarized as follows.

%%%%%%%%%%%%%%%%%%below TW for MBL from paper 2
When $u_0 \in [0, u_I)$, it is proved that there is a constant $\tau_*$ such that for all $\tau \in [0, \tau_*]$, there exists a unique solution of (\ref{eqn:rnereorder2}) connecting $u_+ = u_\alpha$ and $u_- = u_0 $, where $u_\alpha$ is the unique root of the equation
\begin{align}
	G'(u) = \frac{G(u) - G(u_0)} {u - u_0}. \nonumber
\end{align}
When $\tau > \tau_*$, there exists a unique constant $\bar{u} > u_\alpha$, such that (\ref{eqn:rnereorder2}) has a unique solution connecting $u_+ = \bar{u}$ and $u_- = u_0$. For $u_- = u_0 < u_+ = u_B < \bar{u}(\tau)$, the solution of (\ref{eqn:rnereorder2}) will exist only if $u_B \in (u_0, \underline{u})$, where $\underline{u}$ is the unique root in the interval $(u_0, \bar{u}$) of
\begin{align}
	%\label{eqn:uunderline}
	\frac{G(u) - G(u_0)}{u - u_0} = \frac{G(\bar{u}) - G(u_0)} { \bar{u} - u_0}.\nonumber
\end{align}
When $\tau > \tau_*$ and $u_B \in (\underline{u}, \bar{u})$, there is no TW solution of (\ref{eqn:rnereorder2}) connecting $u_+ = u_B$ and $u_- = u_0$. In this situation, the solution profile is non-monotonic, two TWs are used in succession: one from $u_+ = u_B$ to $u_- = \bar{u}$ and one from $u_+ = \bar{u}$ to $u_- = u_0$. For any $u_B \in (\underline{u}, \bar{u})$ and $\tau > \tau_*$, there exists a unique solution of (\ref{eqn:rnereorder2}) such that $u_+ = u_B$, $u_- = \bar{u}$.

\begin{figure}[!htbp]
	\begin{center}
		{\includegraphics[width=2.5in] {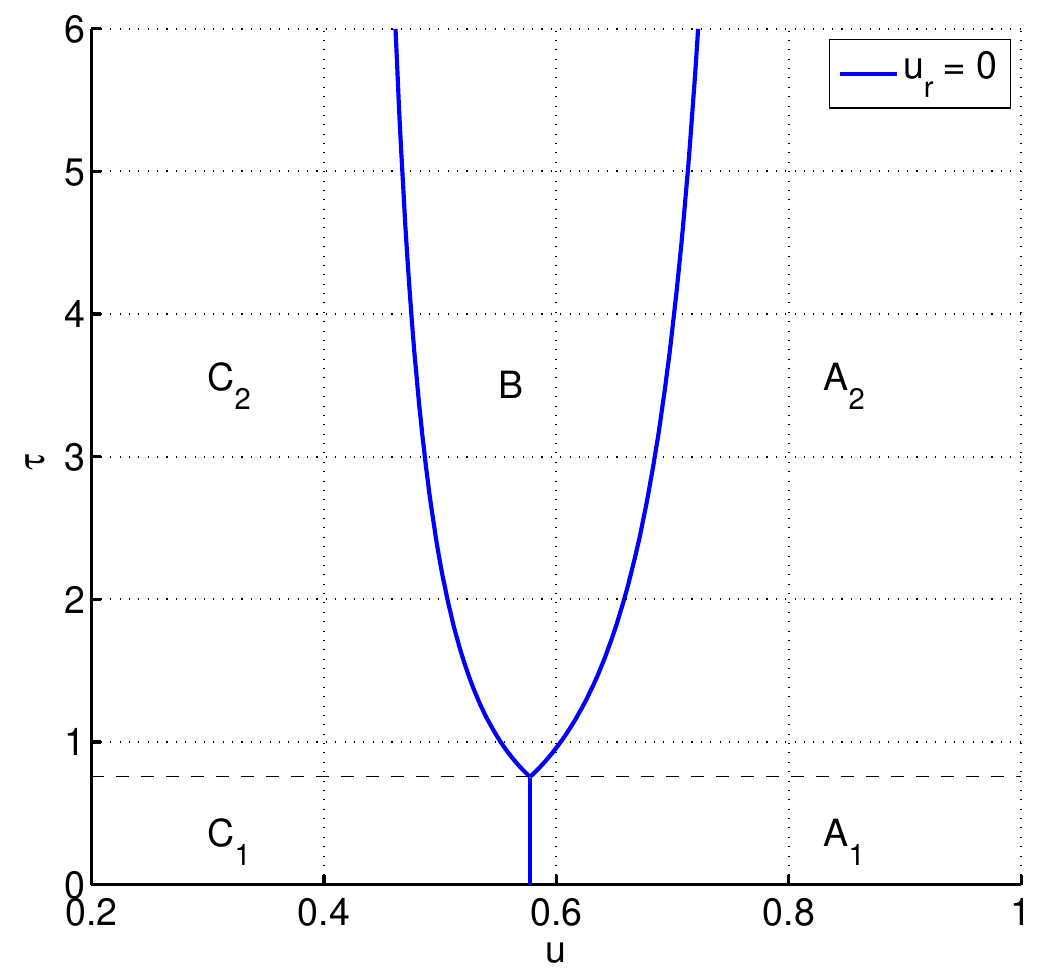}}
		\caption{Bifurcation diagram for options (\ref{eqn:mblbifu}) with $u_0 = 0$.\label{fig:kaobifurcationregions} }
	\end{center}
\end{figure}

For a given $\bar{u} > u_\alpha$, an algorithm to determine the value of $\tau$ is presented in Ref. \cite{van2007new}. This is based on the following concept, invert the function $u(\eta)$ and define the new dependent variable $w(u)= - u'(\eta(u))$, which satisfies
\begin{align}
	%\label{eqn:slope}
	s \tau H(u)  w w' - D(u) w = s ( u - u_-) - [G(u) - G(u_-)], \nonumber
\end{align}
with boundary condition
\begin{align}
	%\label{eqn:slopeboundary}
	w(u_- = u_0) = w(u_+ =  \bar{u}) = 0. \nonumber
\end{align}
The value of $\tau$ corresponding to a given $\bar{u}$ can be computed using a shooting method proposed by \cite{van2007new}. To show the relationship between $\tau \text{-} \bar{u}$, we take $M = 0.5$, $\epsilon = 10^{-3}$, and plot the bifurcation diagram for $u_0 = 0$ in Fig. \ref{fig:kaobifurcationregions}.

When $u_0 < u_I$ and $u_B > u_0$, the traveling solutions can be classified using the five regions in the bifurcation diagram. The results summarized from Ref. \cite{van2007new} are given in Table \ref{tab:kaobifurcationregions}.

\begin{table}
	\caption{Results summarized from Ref. \cite{van2007new}\label{tab:kaobifurcationregions}.}
	\center
	\begin{tabular}{|c|p{300pt}|}\hline
		Region & Solution description \\ \hline
		$(u_B, \tau) \in A_1$ & Rarefaction wave from $u_B$ down to $u_\alpha$ trailing an admissible Lax shock from $u_\alpha$ down to $u_0$ \\ \hline
		$(u_B, \tau) \in A_2$ & Rarefaction wave from $u_B$ down to $\bar{u}$ trailing an undercompressive shock from $\bar{u}$ down to $u_0$ \\ \hline
		$(u_B, \tau) \in B$ & An admissible Lax shock from $u_B$ up to $\bar{u}$ (may exhibit oscillations near $u_+ = u_B$) trailing an undercompressive shock from $\bar{u}$ down to $u_0$ \\ \hline
		$(u_B, \tau) \in C_1$ & An admissible Lax shock from $u_B$ down to $u_0$ \\ \hline
		$(u_B, \tau) \in C_2$ & An admissible Lax shock from $u_B$ down to $u_0$ (may exhibit oscillations near $u_+ = u_B$ \\ \hline
	\end{tabular}
\end{table}

\section{An Adaptive moving mesh finite difference method} \label{sec:adaptive}
In this section, we describe the numerical procedure of the moving mesh FD method for solving the non-equilibrium equations. This method is based on the quasi-Lagrangian approach \cite{weizhang2010adaptive} in which we first transform the physical PDE from the physical coordinates $(x,z,t)$ to the computational coordinates $(\xi,\eta, t)$ and then discretize it using an FD scheme in the space direction and an IMEX scheme in the time direction.

%on page A2321 Approximate the transformation quantity $J = x_\xi$ by a fourth order finite difference (FD4),
% copied from T E Lee 2014 A finite difference moving mesh method base on conservation fro moving boundary problems

In the moving mesh situation, the uniform rectangular mesh is always redistributed as a non-rectangular mesh, thus introduces a difficulty to the FD discretization. In order to fix the problem we apply a coordinate transformation to (\ref{eqn:simform}). Let $(x, z)$ and $(\xi, \eta)$ denote the physical and computational coordinates. Without loss of generality, $(x, z)$ is assumed to be in the interval $\Omega_p = [x_{min}, x_{max}] \times [z_{min}, z_{max}]$ and $(\xi, \eta) \in \Omega_c = [0, 1]\times[0, 1]$. A general coordinate transformation is given by
\begin{align}
	x = x(\xi, \eta, t),\quad  z = z(\xi, \eta, t), \quad t \in [0, T], \nonumber
\end{align}
with boundary condition
\begin{align}
\label{eqn:transbc}
x(0, \eta, t) = x_{min}, \quad x(1, \eta, t) = x_{max}, \quad z(\xi, 0, t) = z_{min}, \quad z(\xi, 1, t) = z_{max}.
\end{align}
Using the transformation formulas:
\begin{align}
	%\label{eqn:transform}
	u_x &= \frac{1}{J} [ (u z_\eta)_\xi - (u z_\xi)_\eta],\nonumber\\
	D(u) u_x &= \frac{D(u)}{J} [ (u z_\eta)_\xi - (u z_\xi)_\eta],\nonumber\\
	(D(u) u_x)_x &= \frac{1}{J} \big[\big(\frac{D(u)}{J} (z_\eta^2 u_\xi - z_\xi z_\eta u_\xi) \big)_\xi + \big( \frac{D(u)}{J} (z_\xi^2 u_\eta - z_\xi z_\eta u_\xi)  \big)_\eta \big],\nonumber
\end{align}
where $J = x_\xi z_\eta - x_\eta z_\xi$ is the Jacobian of the coordinate transformation, the physical PDE can be transformed to its Lagrangian form
\begin{equation}
	\begin{aligned}
		%\label{eqn:tranPDE}
		u_t &+ \frac{1}{J}  \big(\underbrace{z_\eta F(u) - x_\eta G(u)}_{\tilde{F}}\big)_\xi + \frac{1}{J} \big(\underbrace{  x_\xi G(u) - y_\xi F(u)}_{\tilde{G}} \big)_\eta \\
		&+ \frac{1}{J} \big[\big(\underbrace{ \frac{D(u)}{J} ( z_\eta^2 u_\xi + x_\eta^2 u_\xi - z_\xi z_\eta u_\eta - x_\xi x_\eta u_\eta)}_{R}\big)_\xi  + \big(\underbrace{ \frac{D(u)}{J} (z_\xi^2 u_\eta + x_\xi^2 u_\eta - z_\xi z_\eta u_\xi - x_\xi x_\eta u_\xi)_\eta }_{S} \big)  \big]  \\
		&- \frac{\tau}{J} \big[ \big(\underbrace{ \frac{H(u)}{J} (z_\eta^2 u_{t\xi} + x_\eta^2 u_{t\xi} - z_\xi z_\eta u_{t \eta} - x_\xi x_\eta u_{t \eta})}_P \big)_\xi \\
		&- \big( \underbrace{ \frac{H(u)}{J} (z_\xi^2 u_{t\eta} + x_\xi^2 u_{t \eta} - z_\xi z_\eta u_{t \xi} - x_\xi x_\eta u_{t \xi} )}_Q \big)_\eta  \big] = 0, \quad (\xi, \eta) \in \Omega_c.
	\end{aligned}\nonumber
\end{equation}
For convenience, the above equation is written in a simpler form:
\begin{align}
	\label{eqn:simplePDE}
	u_t + \frac{1}{J} \tilde{F}(u)_\xi + \frac{1}{J} \tilde{G}(u)_\eta + \frac{1}{J} [ R_\xi + S_\eta] - \frac{\tau}{J} [P_\xi + Q_\eta] = 0.
\end{align}

\subsection{The spatial discretization}
We will solve (\ref{eqn:simplePDE}) in the computational domain with a method of lines approach. The space discretization results in a large system of ODEs containing both stiff and nonstiff parts which is suitable to be integrated using an IMEX method. By treating the nonstiff advection terms $F(u)$ and $G(u)$ explicitly and the stiff terms $R_\xi$, $S_\eta$, $P_\xi$ and $Q_\eta$ implicitly, we can get a nonlinear system of equations. Since the stiff terms contain functions that depend on $u$: $D(u)$ and $H(u)$, hence we linearize the nonlinear terms by approximating them at $t^n$ instead of at $t^{n+2}$. In this way we can fully exploit the advantages of the IEMX method. Let the space steps $\Delta \xi = 1/\mathrm{NX}$, $\Delta \eta = 1/\mathrm{NZ}$, the computational domain $\Omega_c$ can be partitioned into $\mathrm{NX} \times \mathrm{NZ}$ equal sized cells $[\xi_i, \xi_{i+1}] \times[\eta_j, \eta_{j+1}]$, $i = 0, 1, \cdots, \mathrm{NX}-1, j = 0, 1, \cdots, \mathrm{NZ}-1$. Let $\Delta t$ denote the time step size, the discretization of (\ref{eqn:simplePDE}) can be written as
\begin{equation}
	\begin{aligned}
		\label{eqn:fulldisc}
		\frac{u^{n+1} - u^n}{\Delta t} 
		&+ \frac{1}{J^n_{i,j}} [ \frac{\bar{\tilde{F}}^n_{i+1/2,j} - \bar{\tilde{F}}^n_{i-1/2,j}}{\Delta \xi}]  
		+ \frac{1}{J^n_{i,j}} [ \frac{\bar{\tilde{G}}^n_{i,j+1/2} - \bar{\tilde{G}}^n_{i,j-1/2}}{\Delta \eta}] \\ 
		&+ \frac{1}{J^n_{i,j}} [ \frac{{R}^{n+1}_{i+1/2,j} - {R}^{n+1}_{i-1/2,j}}{\Delta \xi}]  
		+ \frac{1}{J^n_{i,j}} [ \frac{{S}^{n+1}_{i,j+1/2} - {S}^{n+1}_{i,j-1/2}}{\Delta \eta}]  \\
		&- \frac{\tau}{J^n_{i,j}} [ \frac{{P}^{n+1}_{i+1/2,j} - {P}^n_{i-1/2,j}}{\Delta \xi}] 
		- \frac{\tau}{J^n_{i,j}} [ \frac{{Q}^{n+1}_{i,j+1/2} - {Q}^{n+1}_{i,j-1/2}}{\Delta \eta}]  = 0,
	\end{aligned}
\end{equation}
where the advection terms are discretized into conservation forms with $\bar{\tilde{F}}$ and $\bar{\tilde{G}}$ are the numerical fluxes in $\xi$-, $\eta$-direction, respectively:
\begin{align}
	%\label{eqn:numflux}
	\bar{\tilde{F}}_{i+1/2,j} = \bar{\tilde{F}}(u_{i+1/2,j}^-, u_{i+1/2,j}^+), \quad 	\bar{\tilde{G}}_{i,j+1/2} = \bar{\tilde{G}}(u_{i,j+1/2}^-, u_{i,j+1/2}^+).\nonumber
\end{align}
Ref. \cite{kissling2015computation} pointed out that in general a solution containing nonclassical waves cannot be approximated by standard schemes which rely almost entirely on the idea of suppressing variation (e.g. monotone or total-variation-diminishing (TVD) / total-variation-bounded (TVB) schemes). Therefore, we employ the central difference scheme
\begin{align}
	\label{eqn:centralflux}
	\bar{\tilde{F}}(u_{i+1/2,j}^-, u_{i+1/2,j}^+) = \bar{\tilde{F}}(u_{i,j}, u_{i+1,j}) =  \frac{1}{2}[ \tilde{F}(u_{i,j}) + \tilde{F}(u_{i+1,j}) ],
\end{align}
and the local Lax-Friedrichs scheme
\begin{align}
	\bar{\tilde{F}}(u_{i+1/2,j}^-, u_{i+1/2,j}^+) = \frac{1}{2} [ \tilde{F}(u_{i+1/2,j}^-) + \tilde{F}(u_{i+1/2,j}^+) - \max |\tilde{F}_u| \cdot (u_{i+1/2,j}^+ - u_{i+1/2,j}^-)],\nonumber
\end{align}
where the third term stabilizes the scheme by adding dissipation and the maximum is taken between $u_{i,j}^-$ and $u_{i,j}^+$. Two approaches are used to give the values of $u_{i,j}^-$ and $u_{i,j}^+$, one is the standard choice (LLF):
\begin{align}\label{eqn:llf}
	u^{-}_{i+\frac{1}{2}, j} = u_{i,j}, \quad u^+_{i+\frac{1}{2}, j} = u_{i+1,j},	
\end{align}
the other adopts a reconstruction using a linear approximation in each cell (LLFR) \cite{zhang2002adaptive}:
\begin{equation}
	\label{eqn:llfr}
	\begin{aligned}
		&u^{-}_{i+\frac{1}{2},j} = u_{i,j} + \frac{\Delta \xi}{2} s_{i,j}, \quad u^{+}_{i+\frac{1}{2}, j} = u_{i+1,j} - \frac{\Delta \xi}{2} s_{i+1,j}, \\
		&s_{i,j} = \big(\mathrm{sign} (s^-_{i,j}) + \mathrm{sign} (s^+_{i,j})\big) \frac{\|s^-_{i,j} s^+_{i,j}\|}{\|s^-_{i,j}\| + \|s^+_{i,j}\|}, \\
		&s^-_{i,j} = \frac{u_{i,j} - u_{i-1,j}}{\Delta \xi}, \quad s^+_{i,j} = \frac{u_{i+1,j} - u_{i,j}}{\Delta \xi}.
	\end{aligned}
\end{equation}
The discretization for $\bar{\tilde{G}}$ is similar to that of $\bar{\tilde{F}}$.
Then we apply the central difference scheme to the diffusion terms, the mixed derivative terms and the coordinate derivatives, for example:
\begin{equation}
	\begin{aligned}
		R_{i+1/2,j}^{n+1} = &\frac{D(u_{i+1/2,j}^n)}{J_{i+1/2,j}^n} 
		[ \big( (z_\eta|_{i+1/2,j}^n)^2 + (x_\eta|_{i+1/2,j}^n)^2 \big) \frac{u^{n+1}_{i+1,j} - u^{n+1}_{i,j}}{\Delta \xi} \\
		\quad \quad &- \big( z_\xi|_{i+1,j}^n z_\eta|_{i+1,j}^n +  z_\xi|_{i+1,j}^n z_\eta|_{i+1,j}^n \big) \frac{u^{n+1}_{i+1,j+1} - u^{n+1}_{i+1,j-1}}{2 \Delta \eta}], \\
		P_{i+1/2,j}^{n+1/2} = &\frac{D(u_{i+1/2,j}^n)}{J_{i+1/2,j}^n} 
		[ \big( (z_\eta|_{i+1/2,j}^n)^2 + (x_\eta|_{i+1/2,j}^n)^2 \big) \frac{(u^{n+1}_{i+1,j}-u^n_{i+1,j}) - (u^{n+1}_{i,j} - u^n_{i,j})}{\Delta t \Delta \xi} \\
		\quad \quad &- \big( z_\xi|_{i+1,j}^n z_\eta|_{i+1,j}^n +  z_\xi|_{i+1,j}^n z_\eta|_{i+1,j}^n \big) \frac{(u^{n+1}_{i+1,j+1}-u^n_{i+1,j+1}) - ( u^{n+1}_{i+1,j-1} - u^n_{i+1,j-1})}{2 \Delta t \Delta \eta}], 
	\end{aligned}\nonumber
\end{equation}
and 
\begin{equation}
	\begin{aligned}
		\label{eqn:central}
		&x_\xi|_{i,j} = \frac{x_{i+1,j} - x_{i-1,j}}{2 \Delta \xi}, \quad &&x_\eta|_{i,j} =  \frac{x_{i,j+1} - x_{i,j-1}}{2 \Delta \eta}, \\
		&x_\xi|_{i,j+\frac{1}{2}} =  \frac{1}{2} [ (x_\xi)_{i,j} + (x_\xi)_{i,j+1}], \quad &&x_\eta|_{i+\frac{1}{2},j } = \frac{1}{2} [ (x_\eta)_{i,j} + (x_\eta)_{i+1,j}].
	\end{aligned}\nonumber
\end{equation}
By making a discretization of the entire equation (\ref{eqn:fulldisc}) in the way as we described above, and bringing the terms that should be approximated at time $t^{n+1}$ to the left-hand side of the equation and the other terms to the right hand side, we arrive at the following system of equations,
\begin{align}
	%\label{eqn:systems}
	A(\bar{u}^n) \bar{u}^{n+1} = b(\bar{u}^{n}).\nonumber
\end{align}
In order to solve this large system of equations, we adopt an iterative method - the Bi-Conjugate Gradient Stabilized (Bi-CGSTAB) method \cite{van1992bi}  which is provided by the package LASPACK \cite{skalickylaspack}. The implementation of the moving mesh FD method is also realized using LASPACK.
\subsection{An MMPDE-based moving mesh strategy}
In the situation of moving mesh methods, in order to achieve high accuracy, the mesh points may be redistributed in many ways according to the choices of the monitor function. A mesh equation is often solved simultaneously with the transformed PDE so as to generate the mesh positions in tandem with the solution, as the Moving Finite Element method of \cite{miller1981moving}, the Moving mesh PDE (MMPDE) approach \cite{huang1994moving} and the parabolic Monge-Ampere approach of \cite{budd2009moving}, etc. In this paper we adopt the MMPDE6 proposed in \cite{huang1994moving} and use a balanced monitor with directional control \cite{van2010balanced}.
%\begin{equation}\label{eqn:mmpde5}
%	\textrm{MMPDE5:} \left\{ \begin{aligned}
%		\dot{x} = \frac{1}{\tau_t}\nabla \cdot (M \nabla x), \\
%		\dot{z} = \frac{1}{\tau_t} \nabla \cdot (M \nabla z),
%	\end{aligned}
%	\right.
%\end{equation}
The MMPDE6 in 2D reads
\begin{equation}\label{eqn:mmpde6}
	\textrm{MMPDE6:}\quad \left\{ \begin{aligned}
		\bar\nabla \cdot \bar\nabla \dot{x} = -\frac{1}{\tau_{x}}\bar\nabla \cdot (\mathbf{M} \bar\nabla x), \\
		\bar\nabla \cdot \bar\nabla \dot{z} = -\frac{1}{\tau_{z}} \bar\nabla \cdot (\mathbf{M} \bar\nabla z),
	\end{aligned}
	\right.\quad 
	\mathbf{M} = \left[ \begin{array}{cc}
			M_1 & 0 \\
			0   & M_2
	\end{array}\right], 
\end{equation}
subject to the boundary condition (\ref{eqn:transbc}), where $\bar\nabla = [\partial / \partial \xi, \partial / \partial \eta]^T$ is the computational gradient, $\mathbf{M}$ is a diagonal matrix monitor function which controls the mesh concentration, $\tau_{x}$ and $\tau_{z}$ are artificial time parameters determining the time-scale over which a mesh converges to steady state. Ref. \cite{huang1994moving} shows that when solved exactly, the mapping given by (\ref{eqn:mmpde6}) is well defined for all time.
As a boundary condition, it is required that the grid points in the corners do not move. Moreover, the boundary grid points can only move along that boundary. In practice, we solve the one-dimensional version of (\ref{eqn:mmpde6}): $\dot{x}_{\xi \xi} = -\frac{1}{\tau_x}(M_1 x_{\xi})_\xi$ for the horizontal boundaries and $\dot{z}_{\eta \eta} = -\frac{1}{\tau_z} (M_2 z_{\eta})_\eta$ for the vertical boundaries.
%Note that for the limit $\tau_t \rightarrow 0$, the right-hand side of both equations goes to zero and we obtain
%\begin{align}\label{eqn:poisson}
%	\nabla \cdot (M \nabla x) = 0, \\
%	\nabla \cdot (M \nabla z) = 0,
%\end{align}
% A van Dam 2010.Balanced monitoring of flow phenomena in moving mesh methods.
%The key to a successful moving mesh method is a proper monitor function. Firstly, it should detect the relevant flow features, thereby reducing errors caused by the physical flow solver.

\subsection{An adaptive monitor function with directional control}
\begin{figure}[!htbp]
	\label{fig:meshdc}
	\begin{center}
		{\includegraphics[width=2.5in] {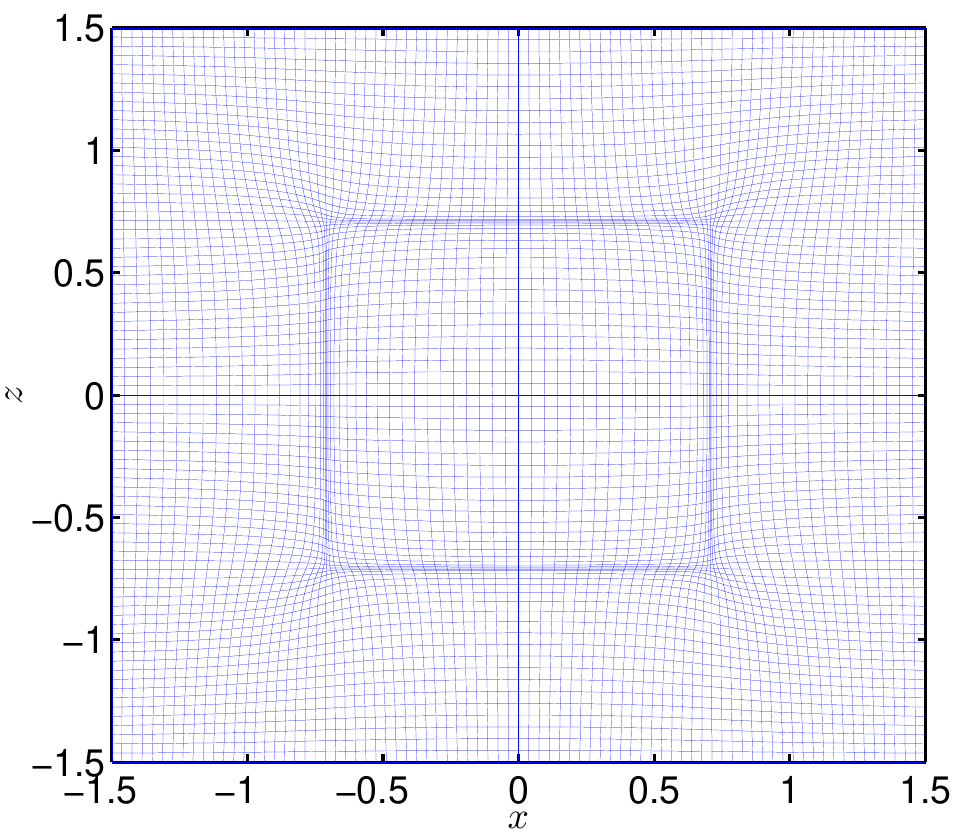}}
		\quad \quad
		{\includegraphics[width=2.5in] {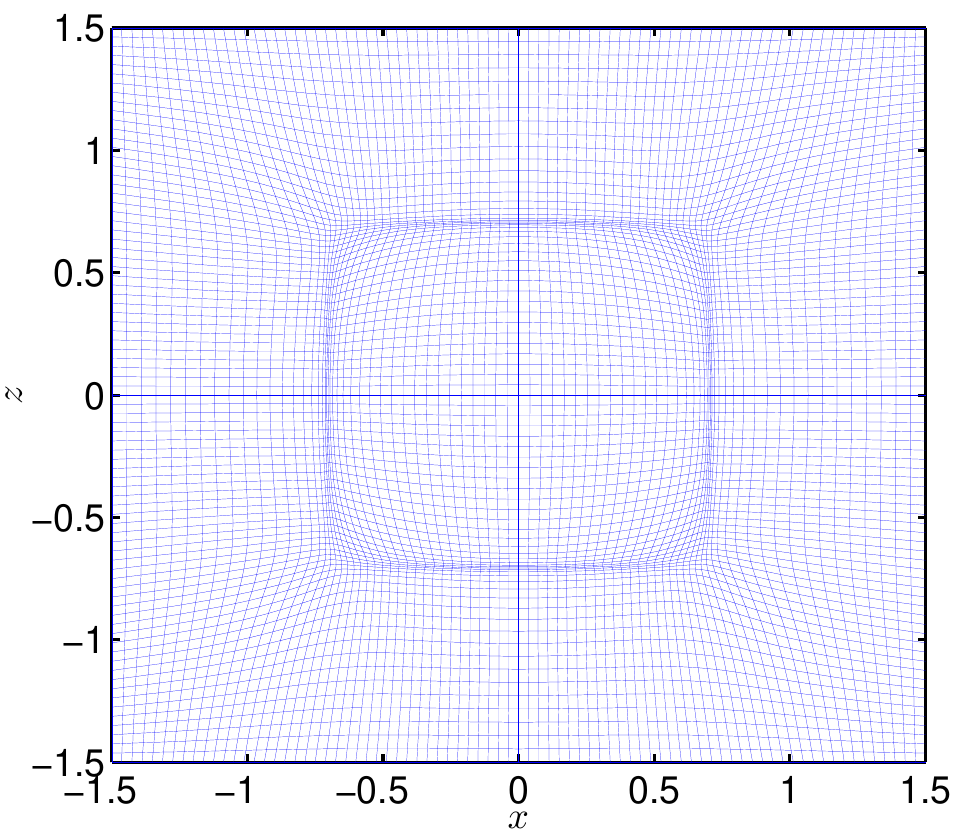}}
		\caption{Initial meshes obtained by time-dependent curvature monitor with directional control (left) and without directional control (right) with parameters $\mathrm{NX} = \mathrm{NZ} = 81, \sigma = 2, \tau_x = \tau_z = 0.1, \kappa = 0.9$. \label{fig:meshdcandnodc}}
	\end{center}
\end{figure}

In the moving mesh method, the monitor function $\mathbf{M}$ is chosen to cluster mesh points at critical regions where more accuracy is needed, thereby reducing errors introduced by the numerical scheme. In this work, we consider a time-dependent monitor function \cite{zegeling2005robust,van2010balanced}
\begin{equation}
		\label{eqn:monitor}
	 M_i = (1 - \kappa) \gamma_i(t) + \kappa \omega_i, \quad i = 1, 2,
\end{equation}
with a normalization for each spatial direction:
\begin{align}
%	\label{eqn:gamma}
	\gamma_i(t) = \int_0^1\!\!\! \int_0^1 \omega_i \mathrm{d} \xi \mathrm{d} \eta. \nonumber
\end{align}
The monitor matrix $\mathbf{M}$ prescribes the monitor values for all directions. If the diagonal elements are identical, for example:
\begin{align}
	%\label{eqn:curvature}
	\omega_i = (|\bar{\Delta} u|^2)^{\frac{1}{4}}, \quad \bar{\Delta} = \frac{\partial^2}{\partial \xi^2} + \frac{\partial^2}{\partial \eta^2}, \nonumber
\end{align}
the mesh adaptation will be nondirectional (isotropic). Ref. \cite{van2010balanced} shows that a directional monitor function could produce much higher quality mesh at negligible costs, thus in this work we impose the directional control (anisotropic). The monitor components $\omega_i$ can be chosen as the arc-length of $u$ in each direction 
\begin{align}
%	\label{eqn:arclength}
	\omega_i = (|\bar{\nabla}_i u|^2)^{\frac{1}{2}}, \quad \bar{\nabla}_1 = \frac{\partial}{\partial \xi}, \bar{\nabla}_2 = \frac{\partial}{\partial \eta}, \nonumber
\end{align}
or the curvature of $u$
\begin{align}
	%\label{eqn:curvature}
	\omega_i = (|\bar{\Delta}_i u|^2)^{\frac{1}{4}}, \quad \bar{\Delta}_1 = \frac{\partial^2}{\partial \xi^2}, \bar{\Delta}_2 = \frac{\partial^2}{\partial \eta^2}. \nonumber
\end{align}
In (\ref{eqn:monitor}), the critical regions are identified by the derivatives computed with respect to the computational coordinates, which are smoother than the physical derivatives. The function $\gamma_i(t)$ averages the derivatives, resulting in a time-dependent monitor function. The ratio of points in the critical regions is denoted by $\kappa$ \cite{huang2001practical}. Thus the monitor matrix $\mathbf{M}$ is a symmetric positive definite matrix with different elements $M_1$ and $M_2$, therefore the mesh adaptivity becomes directional. In Fig. \ref{fig:meshdc} we plot the adaptive meshes obtained using the adaptive curvature monitor function with and without directional control for the initial condition (\ref{eqn:cubeinit}) in Example 3-2. It shows that the monitor function with directional control can identify the critical regions more clearly than the one without directional control.
% from  Tan Zhijun.2004.Moving mesh methods with locally varying time steps

Since the computed monitor components $M_i$, $i = {1,2}$ are usually non-smooth, in order to avoid a very distorted mesh around critical regions, in practice the components are generally smoothed \cite{verwer1988moving,beckett2001numerical,huang1997analysis} before the use for the integration of the MMPDE6.
%In the literature, many smoothing strategies have been proposed \cite{cao1999anr,ceniceros2001efficient,}. Two popular smoothing strategies including the Gaussian filter 
In our computation we apply a smoothing strategy based on a diffusive mechanism in \cite{huang1997analysis}. Similar smoothing strategies have been adopted in \cite{wang2008efficient,hu2011simulating,dong2014adaptive} and obtained good results.
The smoothing equation in \cite{huang1997analysis} is given by
\begin{align}\label{eqn:smoothmonitor}
	[\mathcal{I} - \sigma_s(\sigma_s + 1) \big( (\Delta \xi)^2 \frac{\partial^2}{\partial \xi^2} + (\Delta \eta)^2 \frac{\partial^2}{\partial \eta^2}\big) ] \tilde{M}_i = M_i, \quad  i = 1, 2,
\end{align}
where $\mathcal{I}$ is the identity operator, $\sigma_s$ is the spatial smoothing parameter. By solving (\ref{eqn:smoothmonitor}) we can obtain a smoother monitor function $\tilde{\mathrm{M}}$ which introduces less singularity to (\ref{eqn:mmpde6}), hence MMPDE6 can be solved more efficiently.	

For solving the MMPDE we use the central difference discretization in the space direction and the Euler Backward integrator in the time direction. The monitor function $\mathbf{M}$ is calculated on beforehand, so that the system of equations resulting from the discretization is linear. This system is again solved using the Bi-CGSTAB method.
%% Numerical experiments %%
\section{Numerical experiments} \label{sec:num}
In this section we present numerical results obtained with the moving mesh FD method described in the previous section for a selection of examples. In all examples the time step used satisfies
\begin{align}
	%\label{eqn:adapttime}
	\Delta t = \mathrm{C} ~\mathrm{min}_{i,j} \left( \frac{x_{i+1,j} - x_{i-1,j}}{2 G'(u_{i,j})}, \frac{z_{i,j+1} - z_{i,j}} {2 F'(u_{i,j})} \right),\nonumber
\end{align}
where $\mathrm{C}$ is called a CFL constant. To reduce the time integration error of the IMEX method, we use a CFL number of $0.2$.

In the following subsections, the moving mesh FD method will be investigated with respect to both accuracy and efficiency. 
\subsection{Numerical convergence}
In this section, numerical experiments will be carried out to demonstrate the effectiveness of the moving mesh FD method.

\noindent\textbf{Example 1.} In the first example, we will solve the $1$D MBLE in $z$-direction with the central difference flux (\ref{eqn:centralflux}), the LLF flux (\ref{eqn:llf}) and the LLFR flux (\ref{eqn:llfr}), then we decide which flux scheme is suitable for the computation of the MBLE. The accuracy and effectiveness of the moving mesh method are illustrated by comparing the numerical solutions obtained using the moving mesh with the solutions of the uniform mesh. This example is a $1$-D version of Example 4.5 in \cite{kao2015fast}.

\begin{figure}[!htbp]
	\begin{center}
		{\includegraphics[width=2.5in] {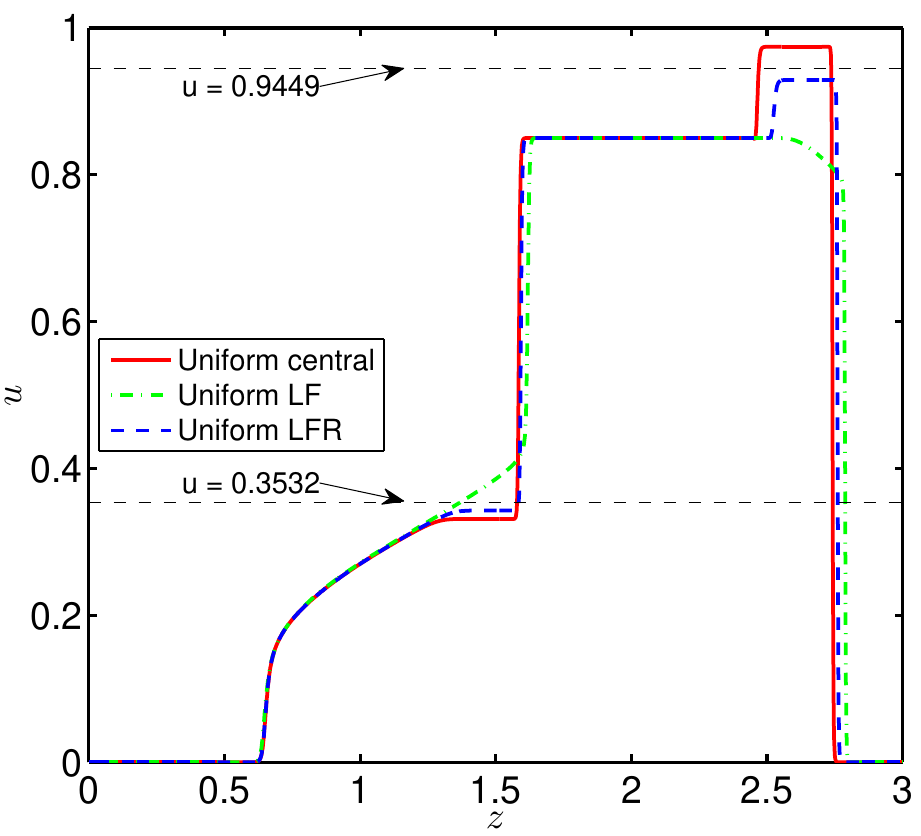}}
		\quad \quad
		{\includegraphics[width=2.5in] {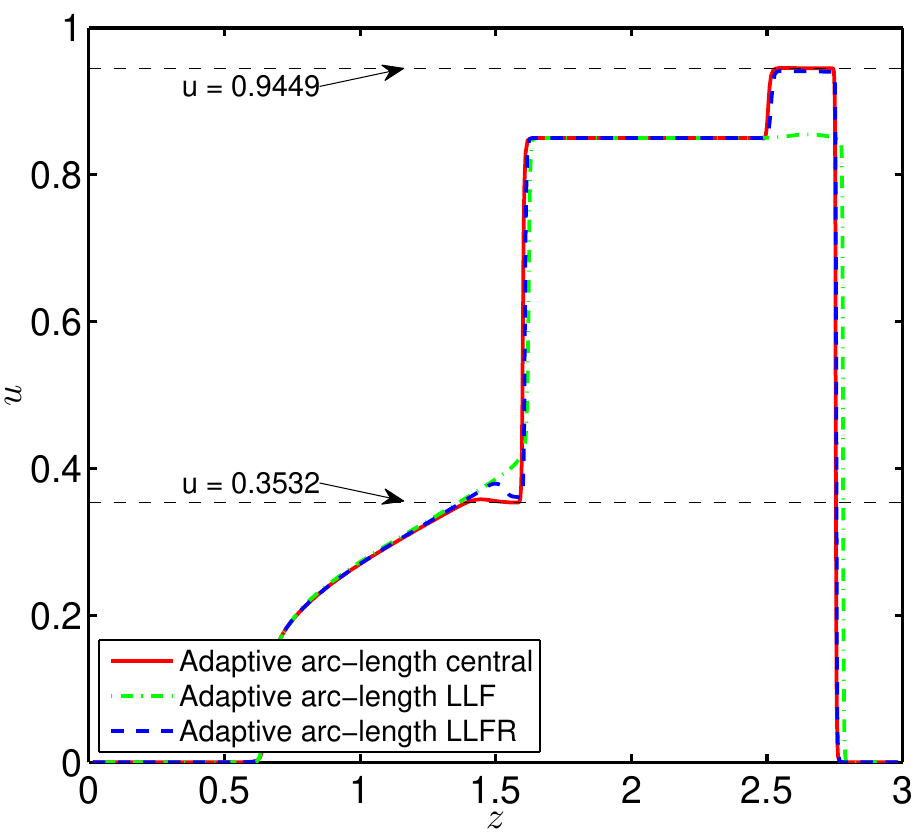}} \\
		{\includegraphics[width=2.5in] {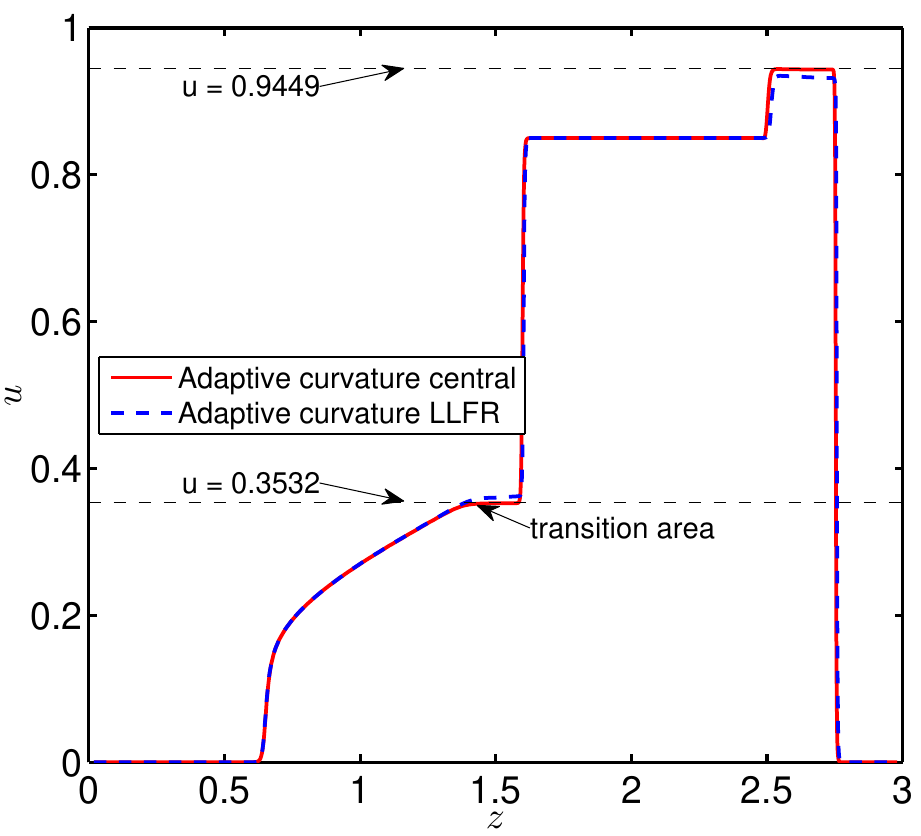}}
		\quad \quad
		{\includegraphics[width=2.5in] {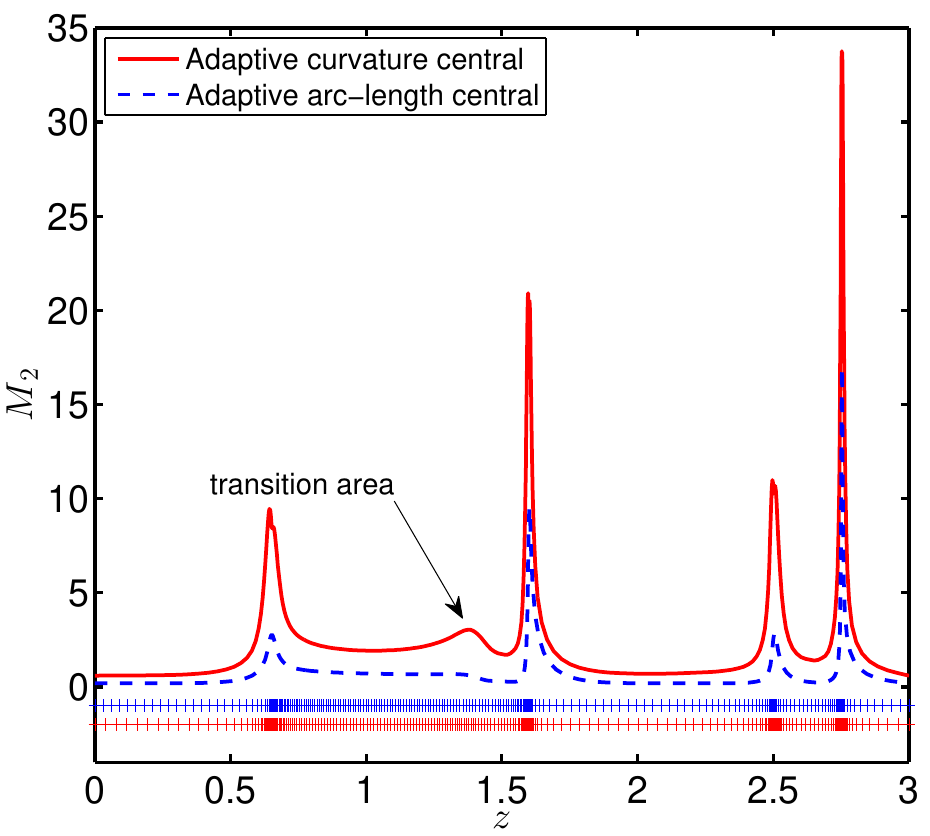}}
		\caption{Solutions computed using the uniform mesh ($\mathrm{NZ} = 2001$, top left) and the moving mesh ($\mathrm{NZ} = 251$, $\sigma = 2, \tau_z = 0.1, \kappa = 0.9$, top right); solutions (bottom left) and monitors (bottom right) obtained by the moving mesh using the arc-length monitor and the curvature monitor. \label{fig:kao4_5_1d}}
% 1d version of Example 4.5 in Kao's paper 
	\end{center}
\end{figure}

In (\ref{eqn:rnere1d}) when $G(u)$, $D(u)$, $H(u)$ are given by
\begin{equation}
	\begin{aligned}
		%\label{eqn:mblf}
		&G(u) = \frac{u^2}{u^2 + M (1 - u)^2}(1 - C(1 - u)^2), \quad D(u) = -\epsilon , \quad H(u) = {\epsilon}^2, \\
		&M = 0.5,  C = 2, \epsilon = 10^{-3}, \tau = 2.5,
	\end{aligned}\nonumber
\end{equation}
with initial and boundary condition
\begin{align}\label{eqn:mblinit}
	u(z,0) = \left\{ \begin{array}{ll}
		0, \quad & z \in [0, 0.75], \\
		0.85,\quad & z\in (0.75, 2.25), \\
		0, \quad  & z \in [2.25, 3],
	\end{array} 	
	\right. \\
	\label{eqn:mblbc}
	u(0,t) = 0, \quad u(3, t) = 0, \quad t \in [0, 0.48].
\end{align}
By applying the shooting method proposed in \cite{van2007new}, we can find that for the initial condition (\ref{eqn:mblinit}), a monotone basin of value $u = 0.3532$ exists in the drainage front together with a non-monotone plateau of value $u = 0.9449$ in the imbibition front. In Fig. \ref{fig:kao4_5_1d}, we plot the numerical solutions obtained by both uniform and moving mesh methods with different fluxes and monitors. In the top left figure one can see that when a uniform mesh with $2001^2$ points is used, the central flux gives a higher plateau and a lower basin, the LLF flux results in no plateau and no basin while the LLFR flux obtains the closest plateau and basin values. These different results are deemed to be caused by the oscillation of the central flux, the diffusion of the LLF flux and the less diffusion of the LLFR flux. The top right figure shows that the when the moving mesh method with the arc-length monitor function is used, the solutions obtained by all three fluxes get improved to some extent: the central flux gives the most accurate plateau and basin values, the LLFR flux gives better plateau value but there is oscillation near the basin area, the solution of the LLF flux gets a little improved but the plateau and basin values are still not acceptable because of the large diffusion of the flux. Although the central flux and the LLFR flux can give very accurate plateau and basin values, near the basin regions oscillations still appear because of a lack of grid points. Therefore, in the bottom left figure we show the results computed using the curvature monitor function. It can be seen that when the transition areas near the basin regions are identified by the curvature monitor (see the bottom right figure), the oscillations are removed, thus the basin profiles of central flux and LLFR flux get improved. From the above observations, it can be concluded that when the moving mesh method is used, the curvature monitor with central flux can give the most accurate plateau and basin values, while in uniform mesh situation the LLFR flux gives closest plateau and basin values among the three fluxes.  

\begin{figure}[!htbp]
	\begin{center}
		{\includegraphics[width=2.5in] {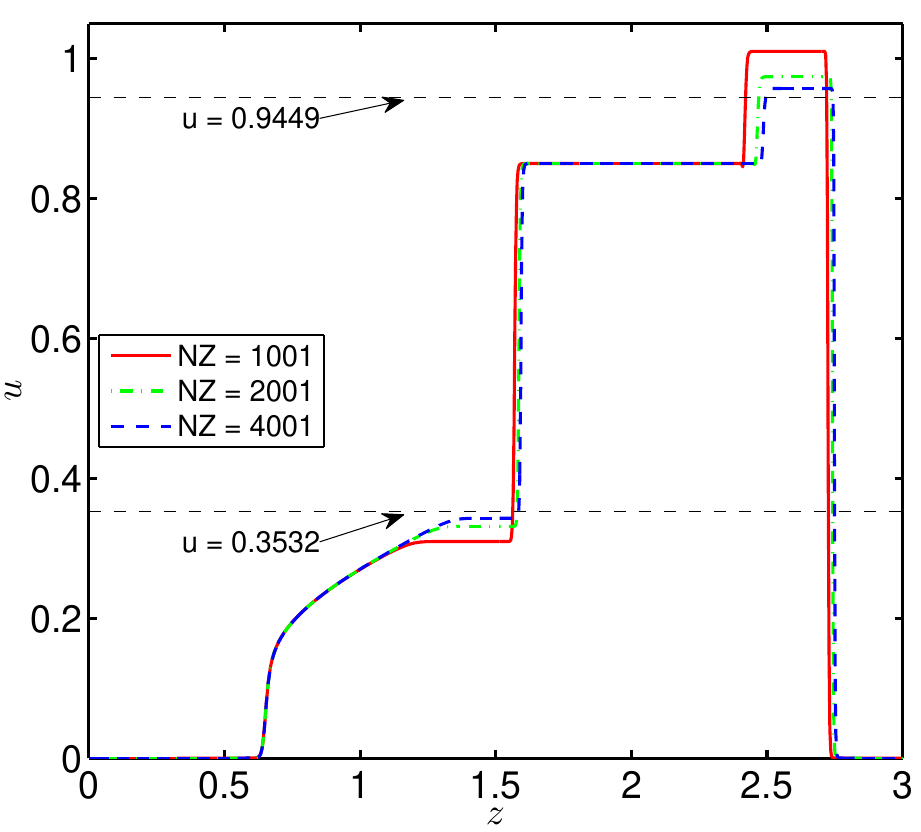}} 
		\quad \quad
		{\includegraphics[width=2.5in] {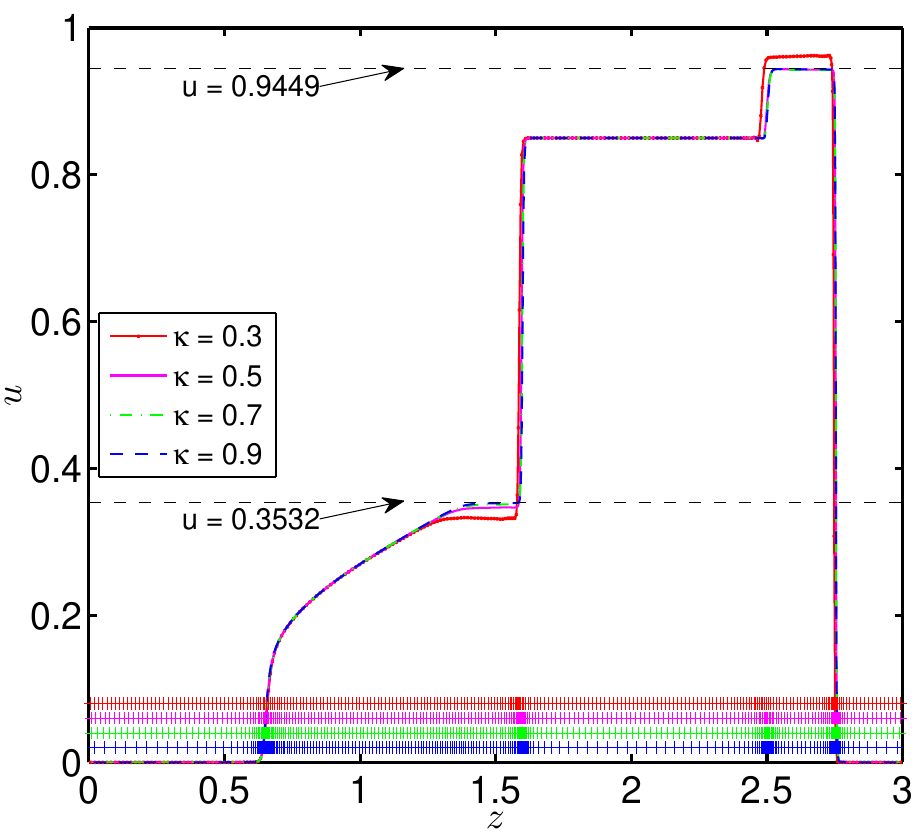}}
		\caption{Example 1: solutions computed at $t = 0.48$ using the uniform mesh (left figure: $\mathrm{NZ} = 1001, 2001, 4001$, central flux) and the moving mesh (right figure: $\mathrm{NZ} = 251$, $\sigma = 2, \tau_z = 0.1$, $\kappa = 0.3, 0.5, 0.7, 0.9$, central flux, curvature monitor).\label{fig:kao4_5_1dconv}}
% 1d version of Example 4.5 in Kao's paper 
	\end{center}
\end{figure}

Fig. \ref{fig:kao4_5_1dconv} shows the convergence of both the uniform method with an increasing number of spatial grid points and the moving mesh method with an increasing of adaptivity parameter $\kappa$. As we can see from the figures, the finer is the uniform mesh, the more accurate solution we get, this obviously shows the numerical convergence of the FD method on uniform meshes. In the moving mesh case, when the adaptivity parameter $\kappa$ becomes larger, more mesh points are clustered at critical regions, which gives plateau and basin heights with more accuracy. It is worth saying that for this example the moving mesh method needs approximately a factor of about 10 fewer grid points than the uniform mesh method to get the same plateau and basin values. The moving mesh method with $251$ points and adaptivity $\kappa = 0.9$ performs even better than the uniform mesh with $4001$ points. Table \ref{tab:comptime} gives a comparison of CPU time between the uniform mesh cases and the moving mesh cases. As we can see, the CPU time increases with increasing mesh size and $\kappa$. The moving mesh of $251$ points with $\kappa = 0.3$ and $\kappa = 0.9$ take almost the same time as the uniform mesh with $1001$ and $2001$ points, respectively, but the moving mesh solutions are more accurate than the uniform mesh solutions.

\begin{figure}[!htbp]
	\begin{center}
		{\includegraphics[width=1.7in] {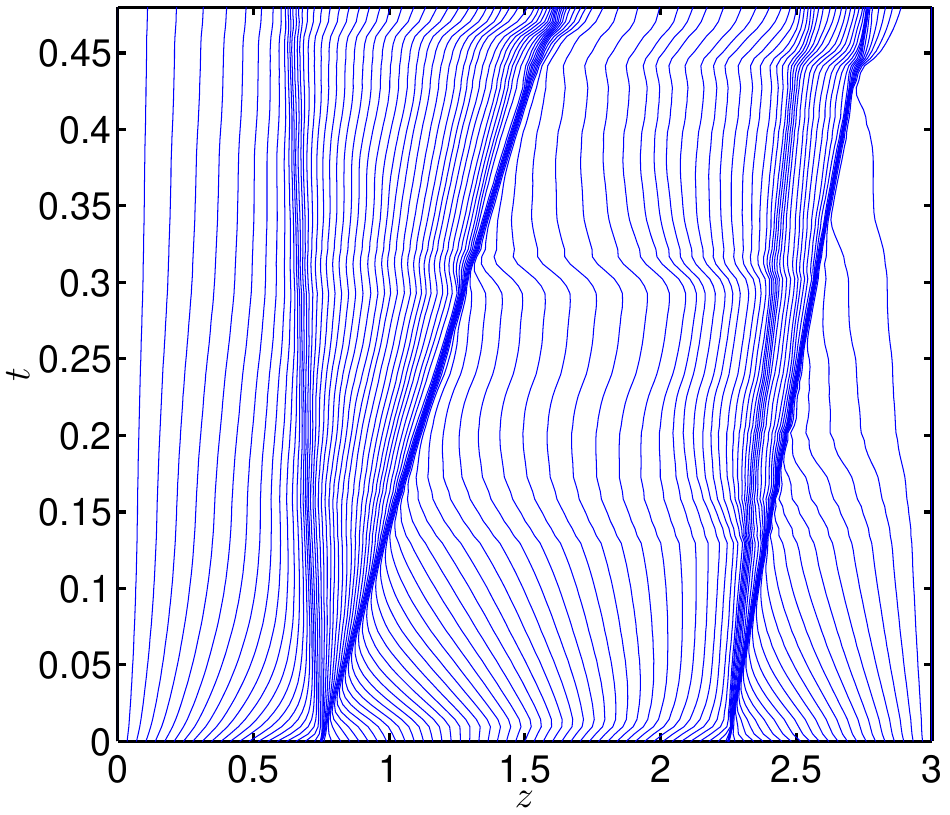}}
		\quad
		{\includegraphics[width=1.7in] {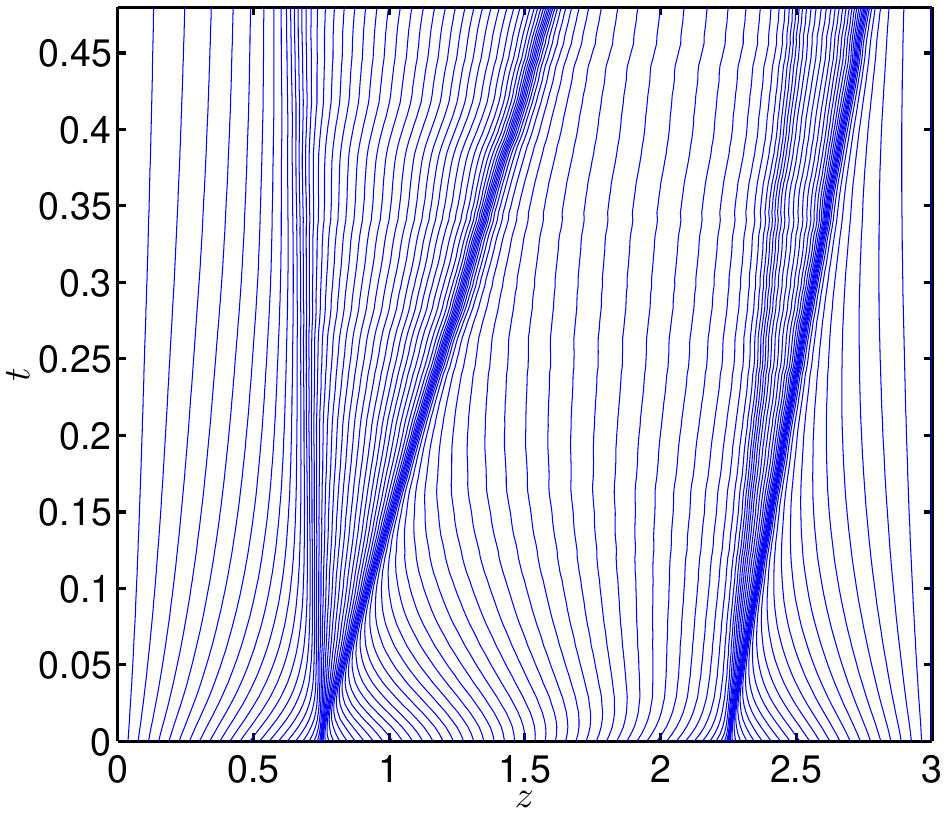}}
		\quad
		{\includegraphics[width=1.7in] {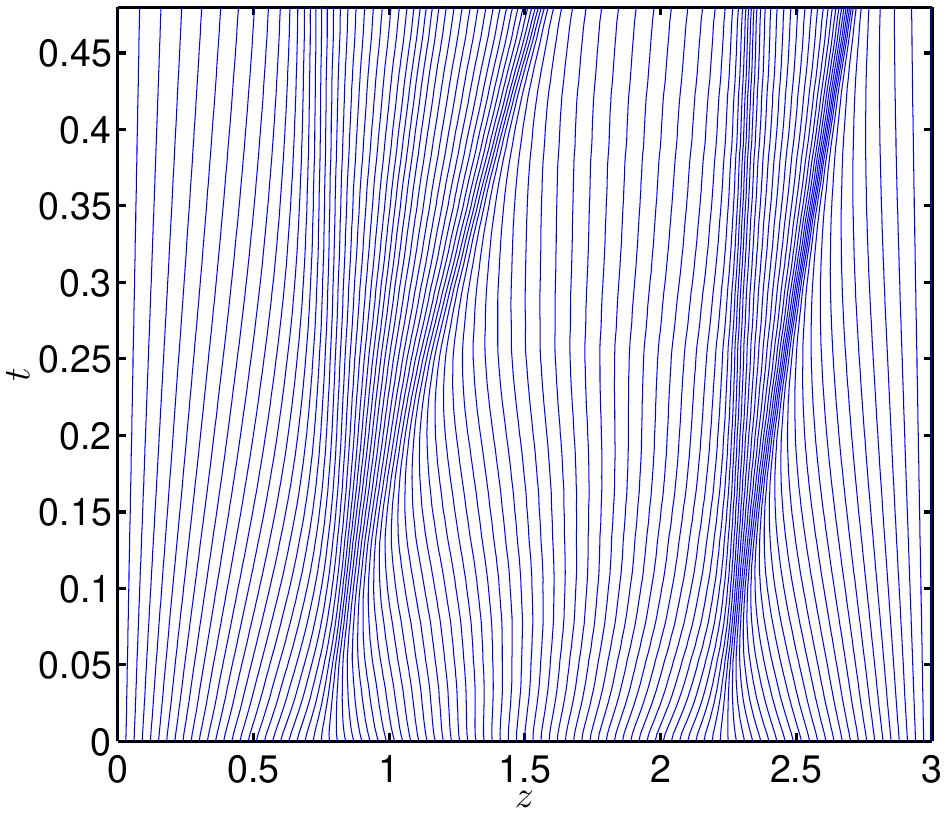}}
		\caption{Grid ($101$ points) trajectories without spatial smoothing (left: $\sigma_s = 0, \tau_s = 0.1$); a grid with smoothing in both space and time variables (middle: $\sigma = 2, \tau_s = 0.1$) and a grid with too much smoothing (right: $\sigma_s = 2, \tau_s = 1$). \label{fig:kao4_5_1d_grid_cmp}}
	\end{center}
\end{figure}

In Fig. (\ref{fig:kao4_5_1d_grid_cmp}) we plot the trajectories of the meshes obtained using different smoothing parameters. When there is non spatial smoothing, the grid trajectories oscillate in the space direction, as the smoothing parameters increase, the grid trajectories become smoother. 
\begin{table}
	\caption{\label{tab:comptime}Comparison of the CPU time $[\mathrm{s}]$ between the uniform mesh and the moving mesh case.}
	\center
	\begin{tabular}{cc|cc}\hline
		\multicolumn{2}{c|}{Uniform mesh}&\multicolumn{2}{c}{Moving mesh}\\
		{Mesh size} & CPU time $[\mathrm{s}]$ & {Adaptivity} & CPU time $[\mathrm{s}]$\\\hline
		1001 & 9.02   & $\kappa = 0.3$ & 9.72 \\
		2001 & 36.93  & $\kappa = 0.5$ & 16.98 \\
		4001 & 164.25 & $\kappa = 0.7$ & 25.91 \\
		{-}  &  -    & $\kappa = 0.9$ & 39.48 \\ \hline
	\end{tabular}
\end{table}

% using central scheme NX = 101

\noindent\textbf{Example 2.} Consider the 1D RNERE (\ref{eqn:rnere1d}) with functions given by $G(u) = u^\alpha, D(u) = \beta u^{\alpha - \beta - 1}$, $H(u) = u^\alpha$ and parameter $\alpha = 3, \beta = 0.25$ with initial and boundary conditions
\begin{equation}
	\left\{
	\begin{aligned}
		%\label{eqn:rnereicbc}
		&u(z,t = 0) = \frac{1}{2}(u_+ - u_-)[1 + \tanh(\frac{100(z - 0.9(z_{max}-z_{min}))}{z_{max} - z_{min}})] + u_-, \\
		&u(z = 0,t) = u^-, u(z = 4,t) = u^+,
	\end{aligned}
	\right.\quad z \in [0, 4]. \nonumber
\end{equation}

\begin{figure}[!htbp]
	\begin{center}
		{\includegraphics[width=2.5in] {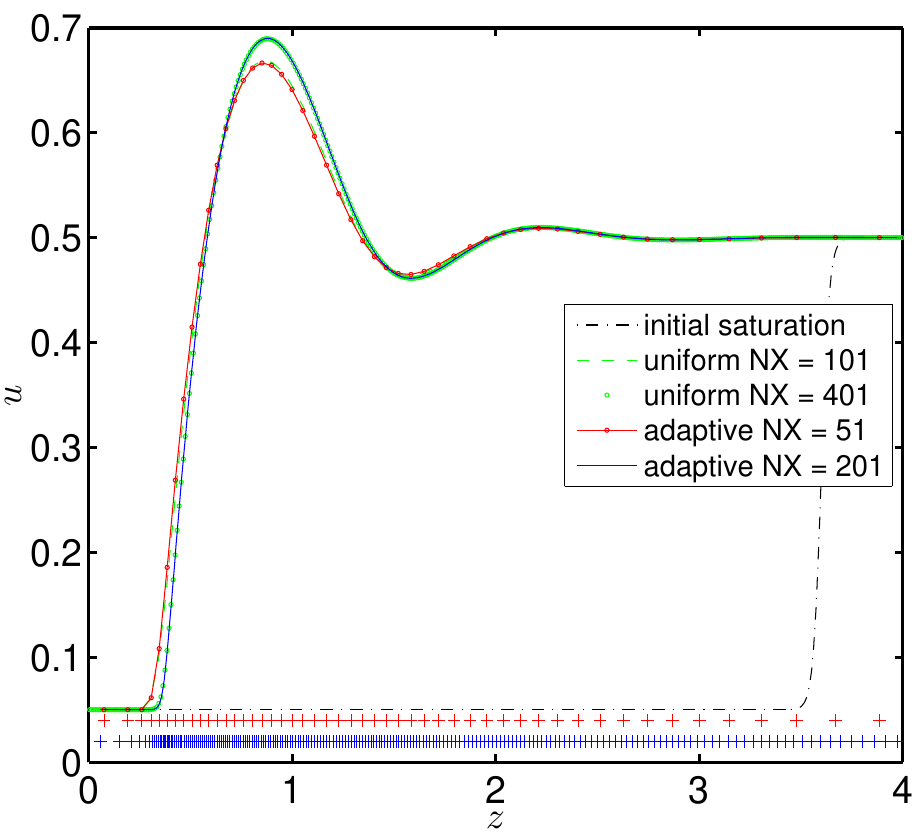}}
		\quad \quad
		{\includegraphics[width=2.5in] {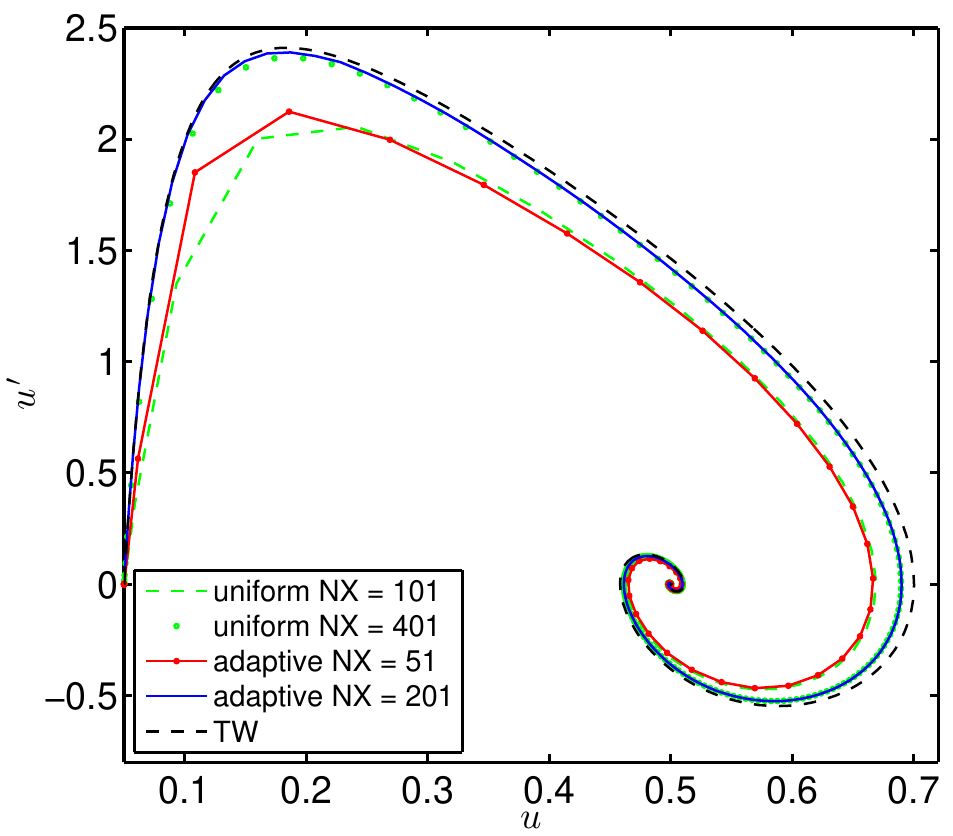}}
		\caption{Solutions (left) and phase planes (right) computed at $t = 12$ using the moving mesh ($\mathrm{NZ} = 51, 201$, $\sigma = 2, \tau_z = 0.1, \kappa = 0.9$) and uniform mesh ($\mathrm{NZ} = 101, 401$). \label{fig:rnere1d_cmp}}
	\end{center}
\end{figure}

Fig. \ref{fig:rnere1d_cmp} presents the initial condition together with the solutions profiles and phase planes computed using the uniform mesh and moving mesh at time $t = 12$. We choose $\mathrm{NZ} = 101, 401$ for the uniform mesh and $\mathrm{NZ} = 51, 201$ for the moving mesh. It shows that in the moving mesh situation, the mesh points are clustered near the critical regions, which helps to improve the accuracy of the solutions. The plots of the phase planes also show that when the meshes are refined, both the uniform and moving mesh profiles converge to the TW result. It is worth saying that with the moving mesh method, the solutions computed using $NZ = 201$ points is comparable with the uniform mesh solution using $401$ points. When $\mathrm{NZ} = 200$, the moving mesh solution almost coincides with the TW solution.

\subsection{Numerical experiments in 2D}
%In this section, we will carry out some numerical experiments to demonstrate the performance of the moving mesh method.
\noindent\textbf{Example 3-1.} The first 2D problem is concerned with the MBLE (\ref{eqn:simform}) without dynamic capillary pressure and the functions are given by
\begin{align}
	%\label{eqn:fluxf}
	& F(u) =  
		\begin{aligned}
			\frac{u^2}{u^2 + ( 1- u)^2)},\\
		\end{aligned}
		\nonumber \\
	& G(u) = f(u) (1 - 5(1 - u)^2),\nonumber \\
	& D(u) = 0.01, \quad H(u) = 0.01^2, \nonumber
\end{align}
the initial data is
\begin{equation}
	u(x, z, 0) = \left\{
		\begin{aligned}
			1, \quad & x^2 + z^2 < 0.5, \\
			0, \quad &\mathrm{otherwise},
		\end{aligned}
\right. \nonumber
\end{equation}
considered in the square domain $[-1.5, 1.5] \times [-1.5, 1.5]$.

\begin{figure}[!htbp]
	\begin{center}
		{\includegraphics[width=1.4in] {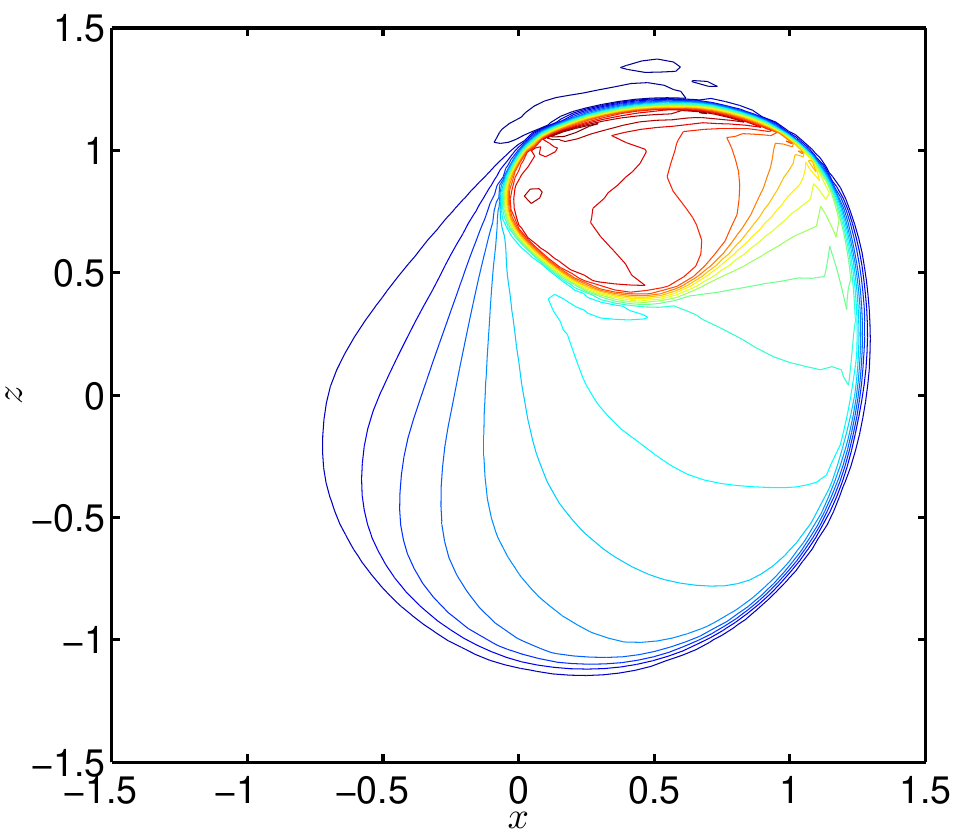}}
		{\includegraphics[width=1.4in] {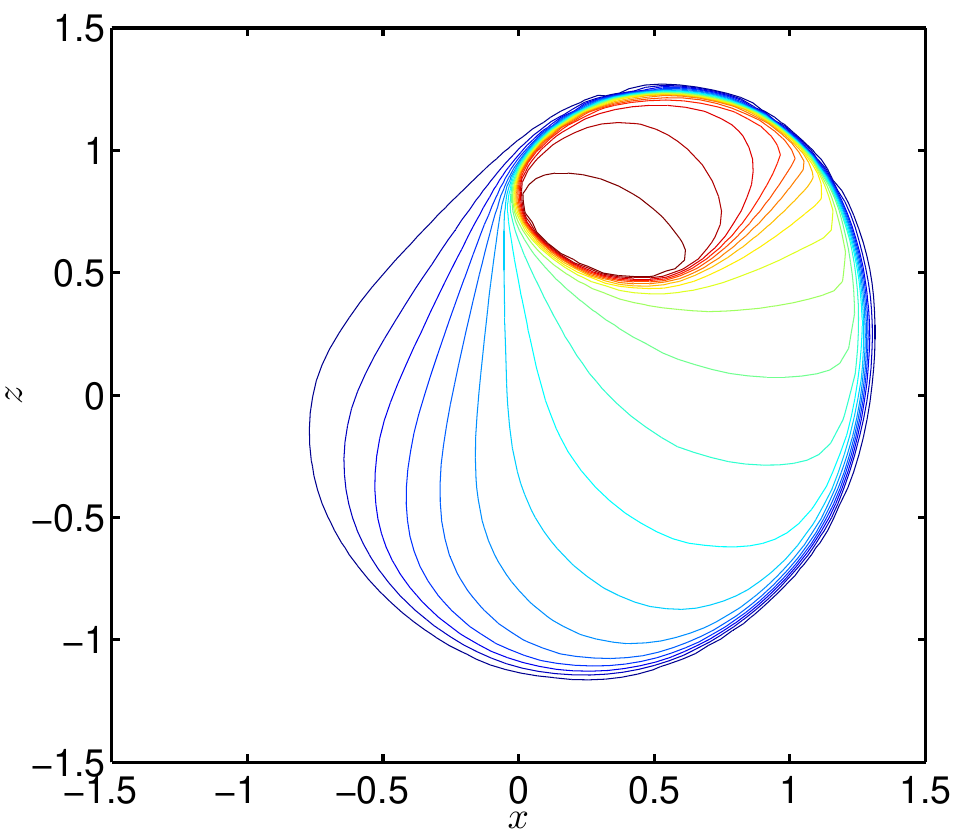}}
		{\includegraphics[width=1.4in] {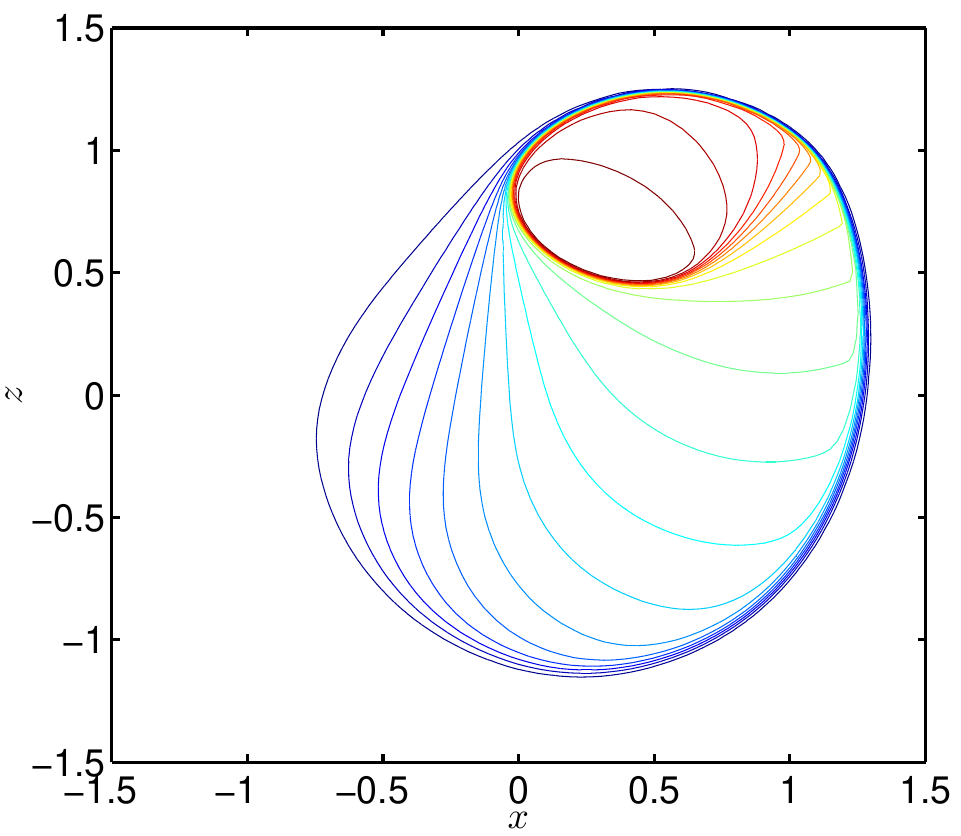}}
		{\includegraphics[width=1.4in] {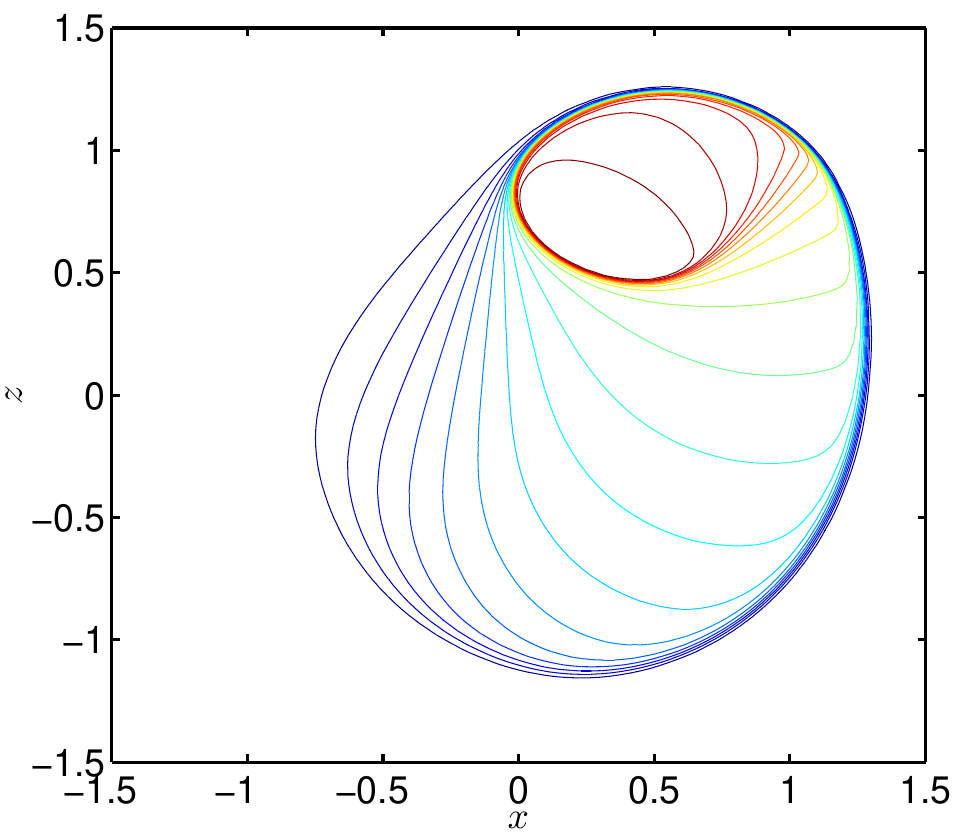}} \\
		{\includegraphics[width=1.4in] {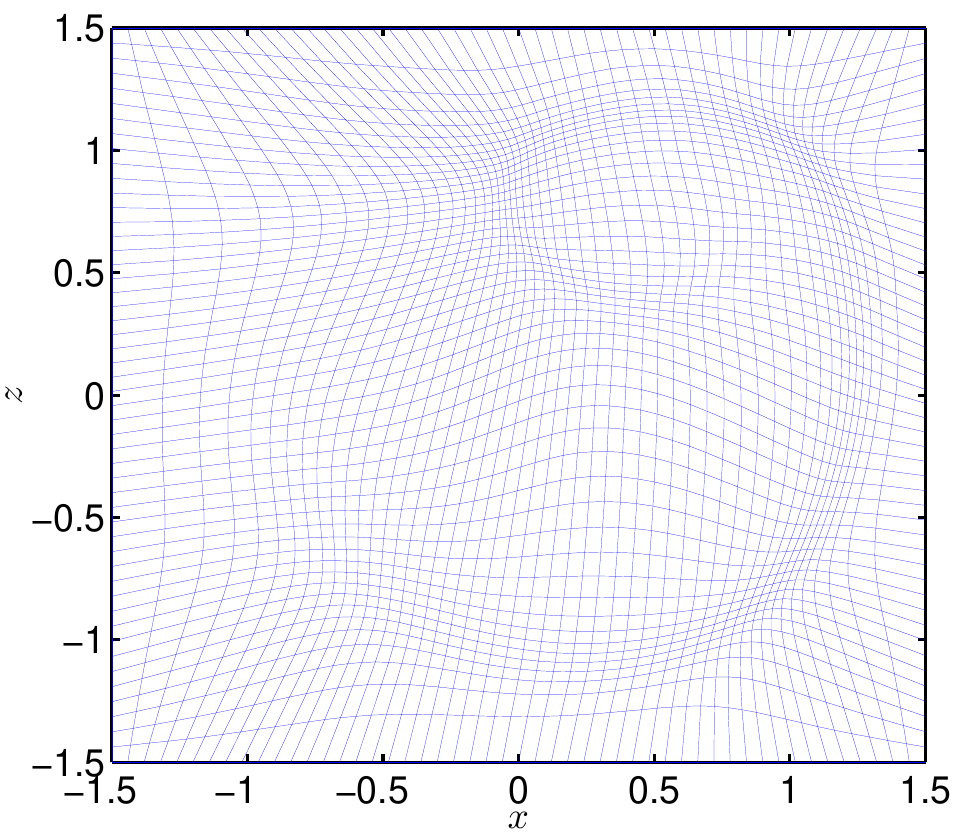}}
		{\includegraphics[width=1.4in] {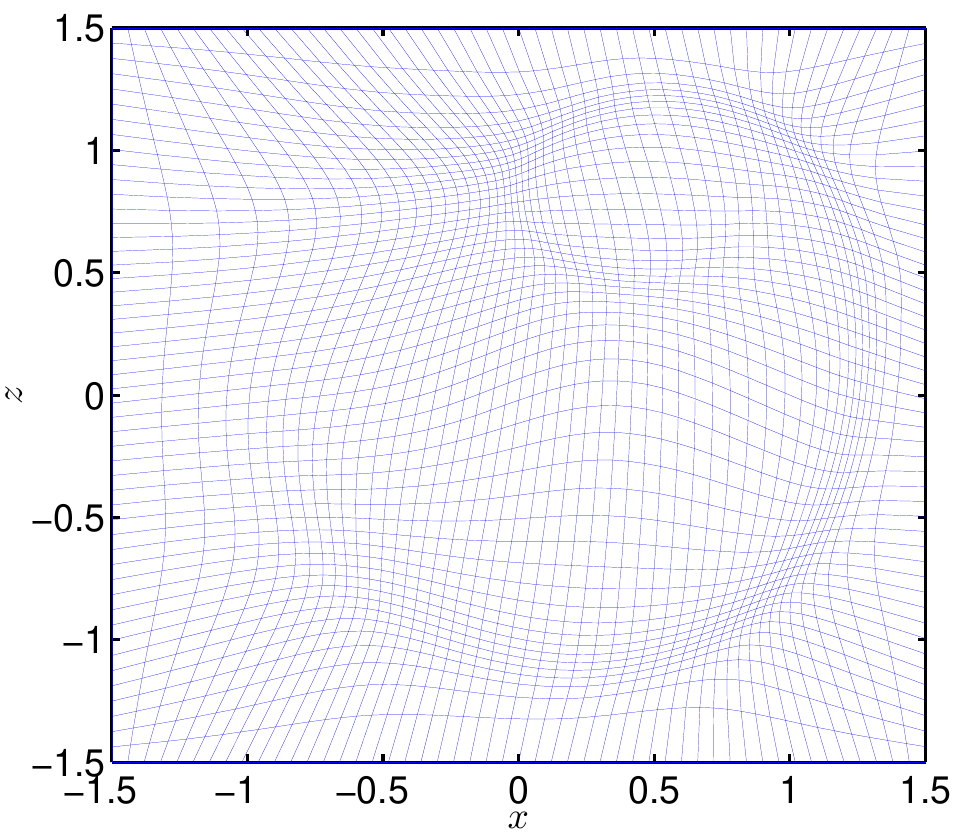}}
		{\includegraphics[width=1.4in] {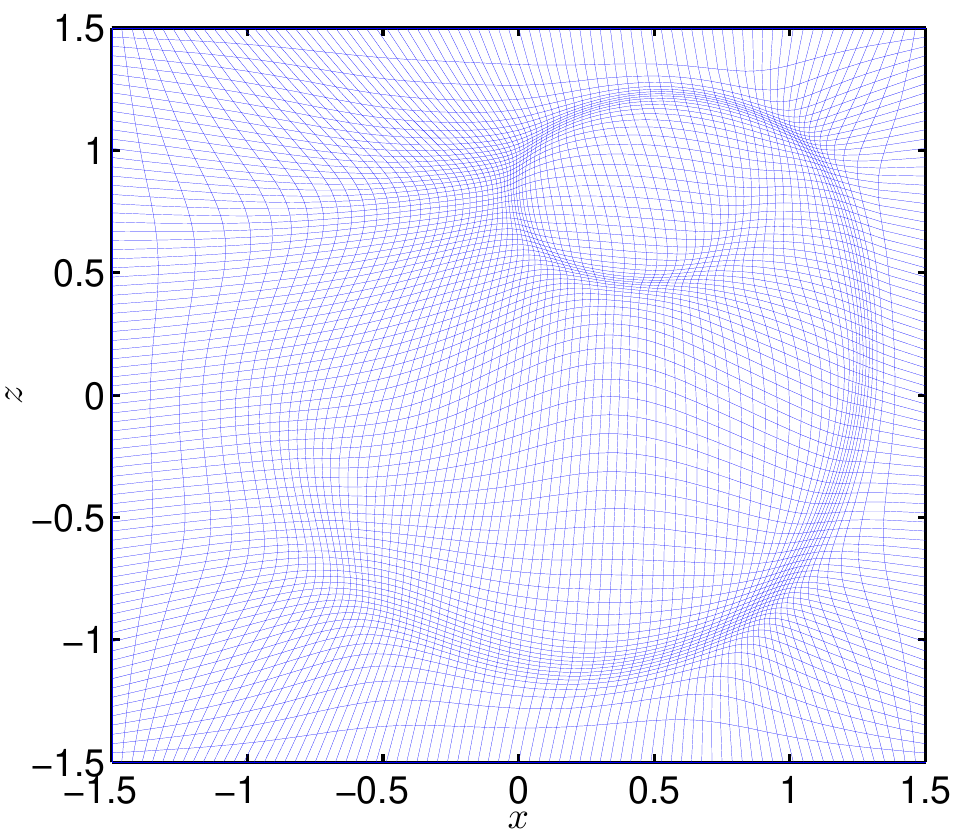}} 
		{\includegraphics[width=1.4in] {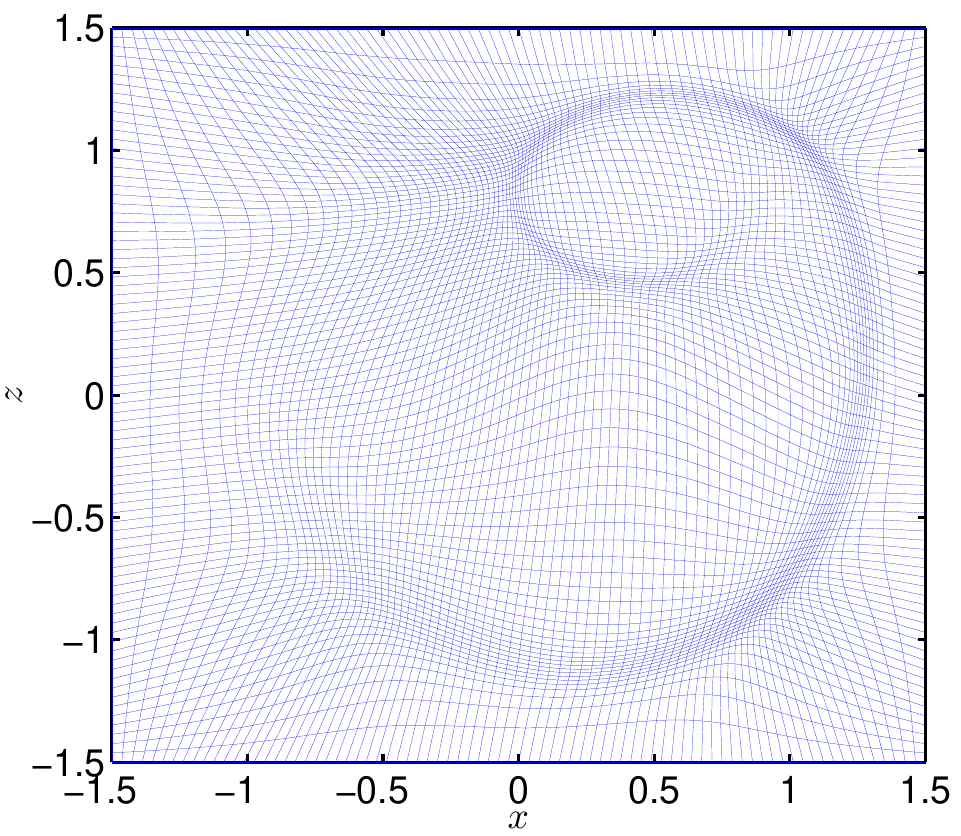}}
		\caption{Example 3-1 with $\tau = 0$: adaptive mesh solutions (top, $\sigma = 2, \tau_x = \tau_z = 0.1$, $\kappa = 0.9$) and corresponding meshes (bottom) at $t = 0.5$. From left to right: central flux $51^2$ points, LLFR flux $51^2$ points, central flux $81^2$ points, LLFR flux $81^2$ points. \label{fig:tangtao2dbl}}
	\end{center}
\end{figure}

This example is taken from \cite{karlsen1999corrected} and has no exact solution. Zhang and Tang \cite{zhang2002adaptive} solved this equation with an adaptive moving mesh finite volume method. Their results shows that the adaptive mesh solutions are more accurate than the uniform mesh ones. Since Example 1 demonstrates that the LLF flux is too diffusive, in Fig. \ref{fig:tangtao2dbl} we only present the moving mesh solutions obtained by the central flux and the LLFR flux on different meshes using the curvature monitor. It is observed that on a mesh with $51^2$ points, the central flux will cause oscillations near the upper front, while the LLFR flux gives smoother profiles. However, if we increase the mesh size to $81^2$, there is no oscillation in the central flux solution and the solution is very close to the LLFR solution. By Comparing Fig. (\ref{fig:tangtao2dbl}) with the results presented in Ref. \cite{zhang2002adaptive}, we may draw the conclude that the moving mesh FD method performs as good as the moving mesh finite volume method.

\begin{figure}[!htbp]
	\begin{center}
		{\includegraphics[width=2.5in] {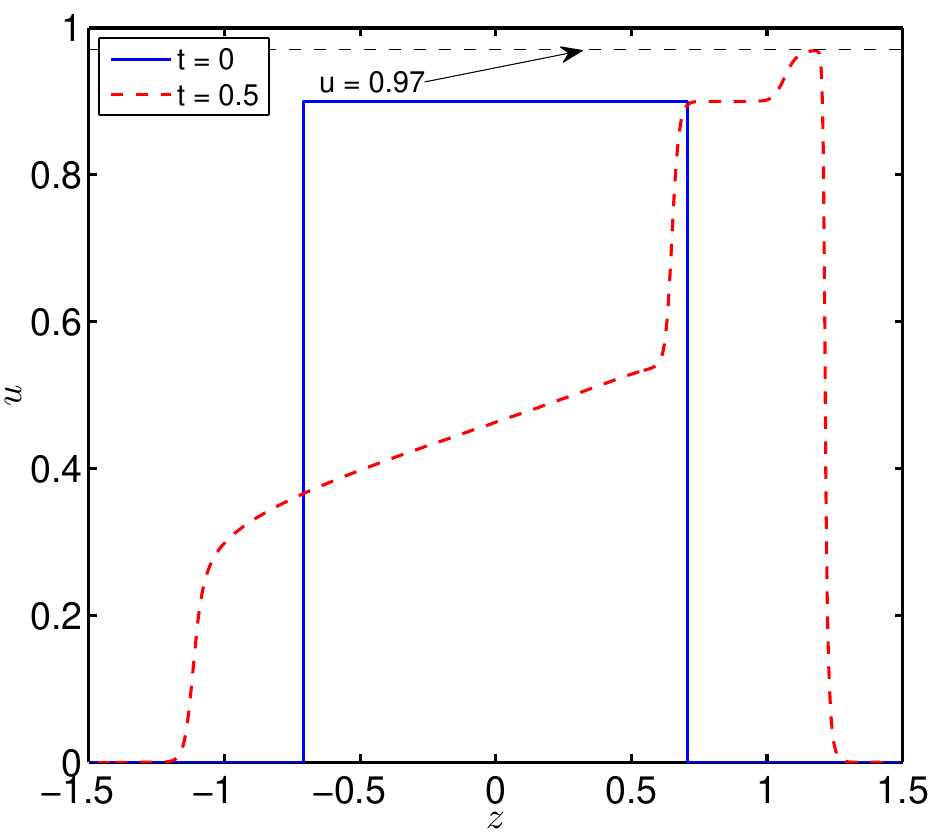}} %central flux
		{\includegraphics[width=2.5in] {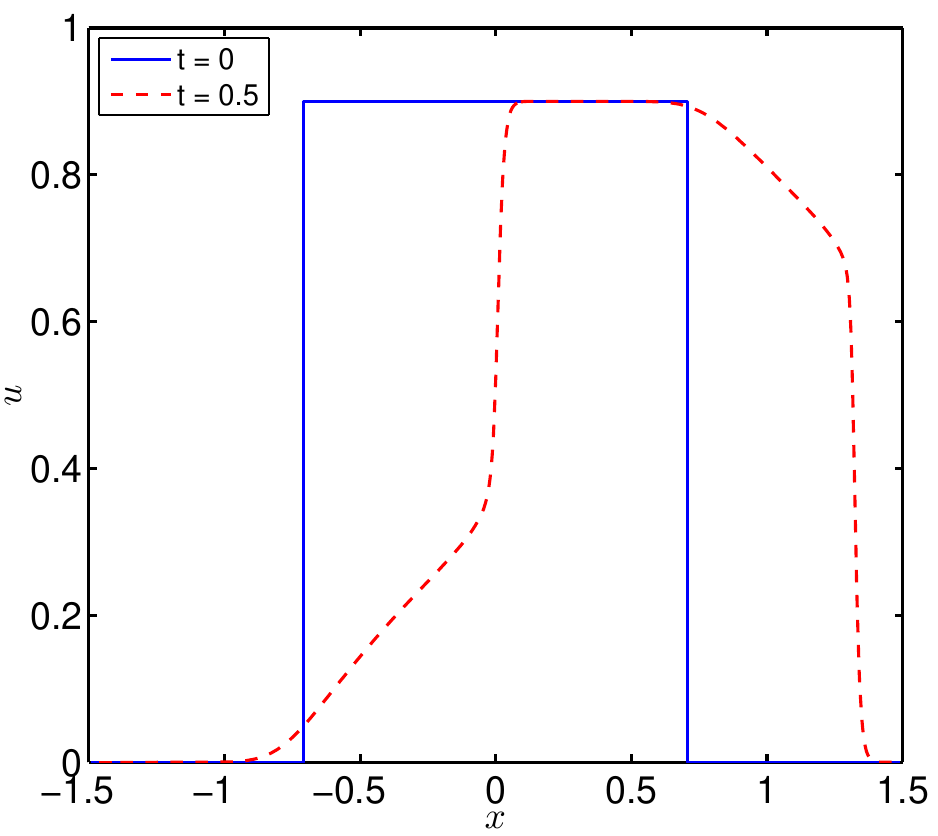}} %central flux
		\caption{1D MBLE with initial condition (\ref{eqn:1dtangtao}) (left) and (\ref{eqn:1dtangtaox}) (right) at $t = 0.5$.
		\label{fig:tangtao1dzaxis}}
	\end{center}
\end{figure}

\noindent\textbf{Example 3-2.} Next, we study two different initial conditions of the 2D MBLE with dynamic capillary pressure term. When $\tau$ is not zero, we can use the TW analysis in Section \ref{sec:twmble} to predict the behavior of the solution. Choosing $\tau = 0.5$ and consider the 1D MBLE in the $z$ direction, if the initial condition is taken as
\begin{equation}
	\label{eqn:1dtangtao}
	u(z, 0) = \left\{
		\begin{aligned}
			0.9, \quad & |z| < \sqrt{0.5}, \\
			0, \quad &\mathrm{otherwise},
		\end{aligned} \quad z \in [-1.5, 1.5],
\right. 
\end{equation}
the TW analysis shows that in the $z$ direction, a saturation plateau of height $u = 0.97$ will appear at the shock front (see Fig. \ref{fig:tangtao1dzaxis} left). For the 1D MBLE in the $x$ direction, taking the initial condition as 
\begin{equation}
	\label{eqn:1dtangtaox}
	u(x, 0) = \left\{
		\begin{aligned}
			0.9, \quad &|x| < \sqrt{0.5}, \\
			0, \quad &\mathrm{otherwise},
		\end{aligned} \quad x \in [-1.5, 1.5],
\right. 
\end{equation}
the TW analysis shows that $\tau$ is too small to produce saturation overshoot, only monotone solution exists in the $x$-direction (see Fig. \ref{fig:tangtao1dzaxis} right). 

\begin{figure}[!htbp]
	\begin{center}
		\subfigure[Plateau height = $1.0253$]
		{\includegraphics[width=1.7in] {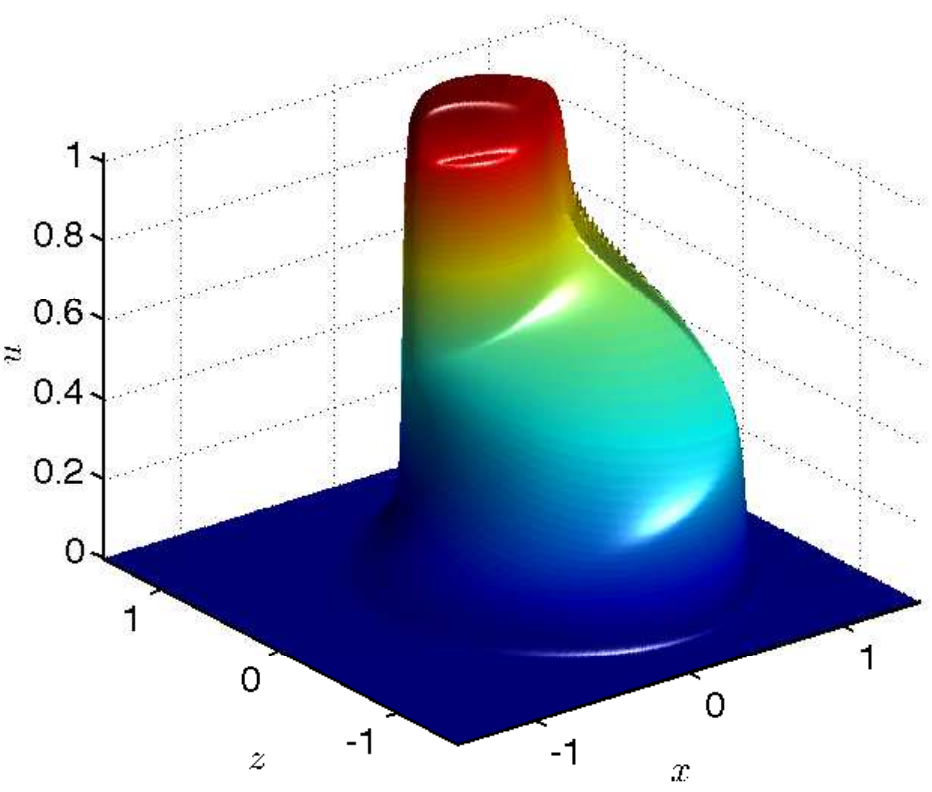}} 
		\subfigure[Plateau height = $0.9698$]
		{\includegraphics[width=1.7in] {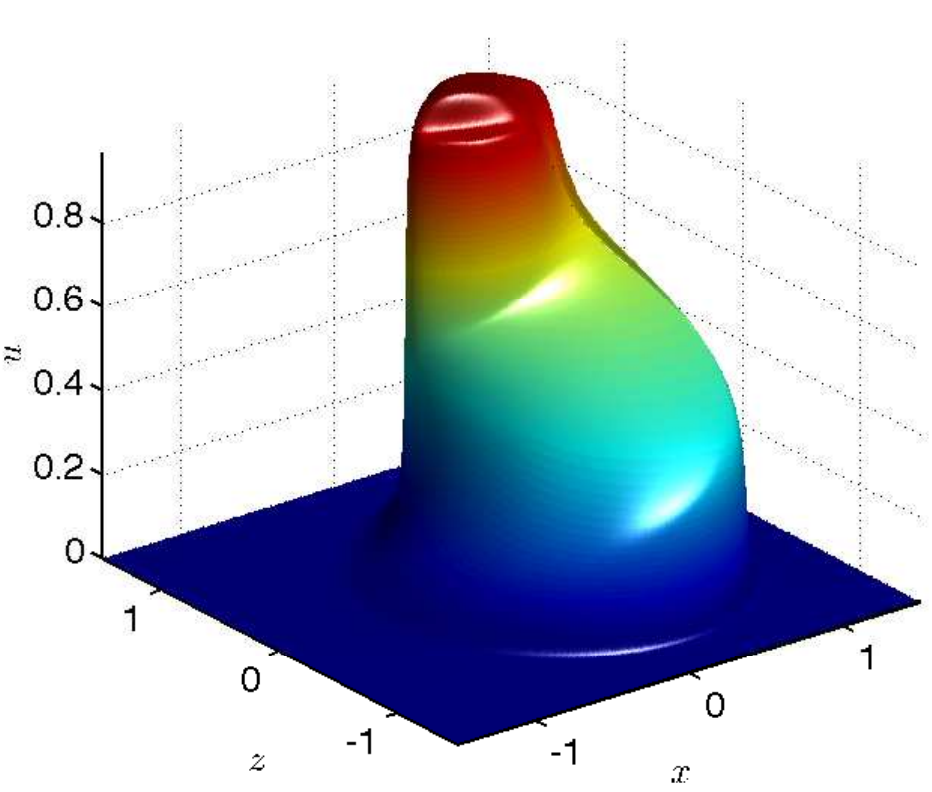}} 
		\subfigure[Plateau height = $0.9859$]
		{\includegraphics[width=1.7in] {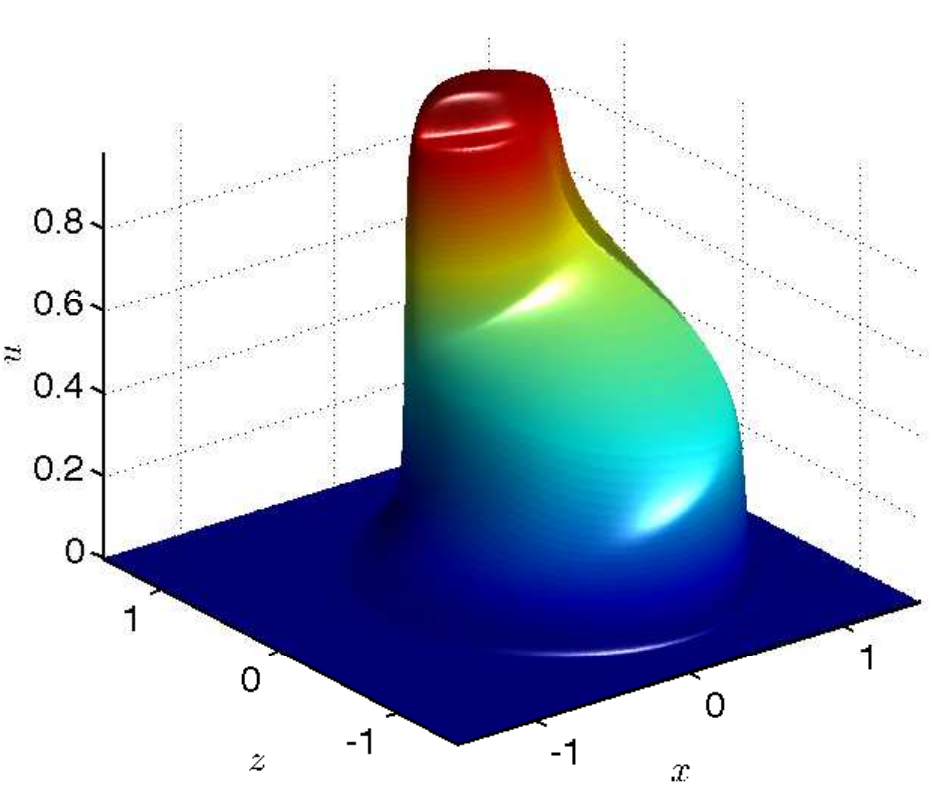}} \\
		\subfigure[Plateau height = $0.9327$]
		{\includegraphics[width=1.7in] {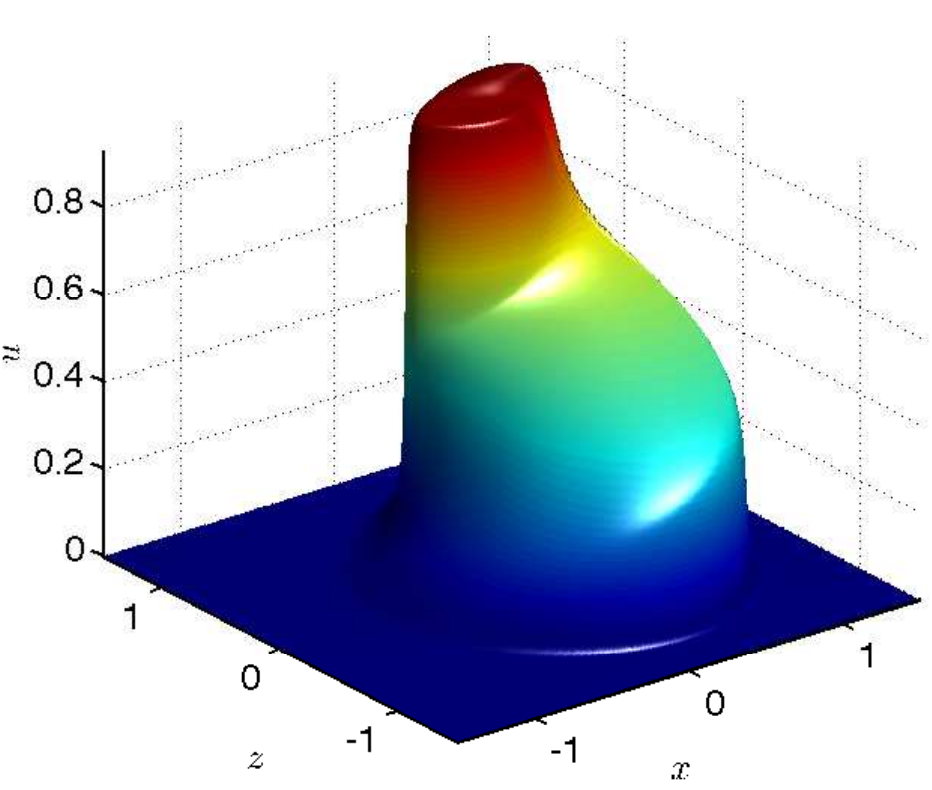}} %LLFR flux
		\subfigure[Plateau height = $0.9571$]
		{\includegraphics[width=1.7in] {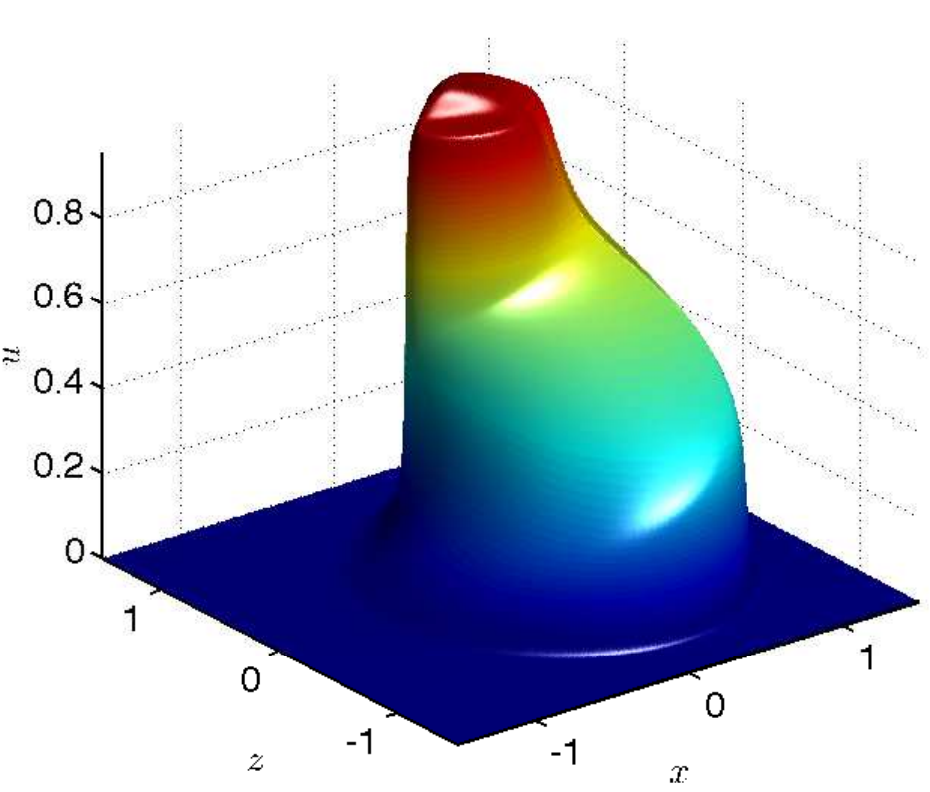}}
		\subfigure[Plateau height = $0.9768$]
		{\includegraphics[width=1.7in] {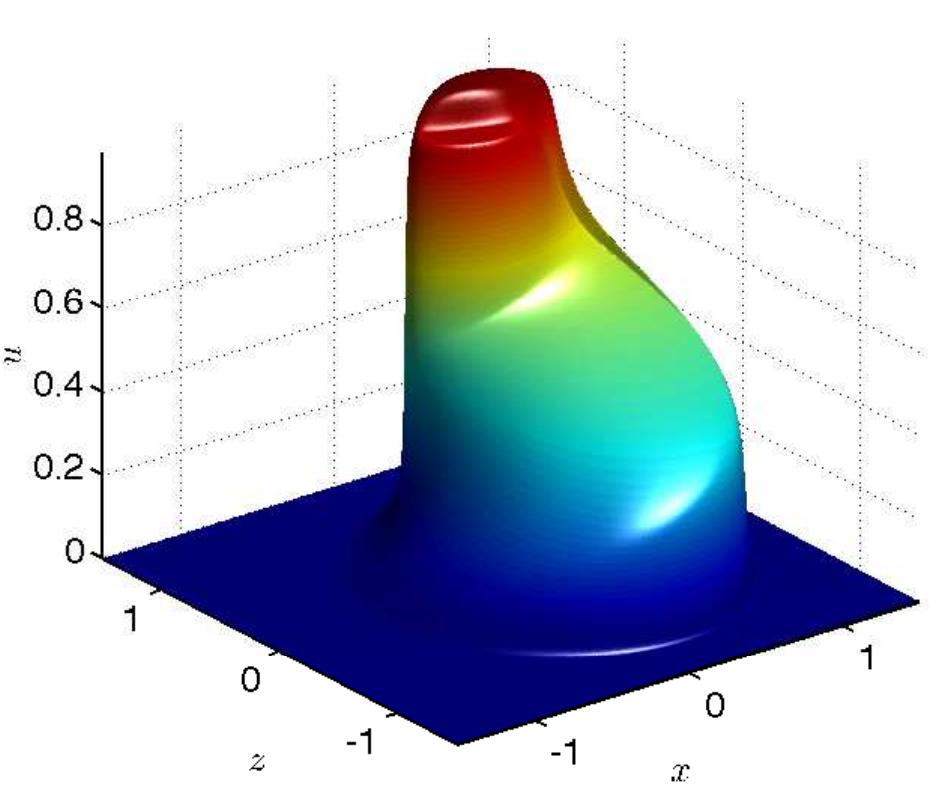}}
		\caption{Example 3-2 with initial condition (\ref{eqn:cyliinit}) at $t = 0.5$: solutions obtained by the uniform mesh and adaptive mesh. Top row: central flux; bottom row: LLFR flux. Left column: uniform mesh $301^2$ points; middle column: moving mesh $301^2$ points ($\sigma = 2, \tau_x = \tau_z = 1, \kappa = 0.6$); right column: uniform mesh $1001^2$ points. \label{fig:tangtao2dmbl}}
	\end{center}
\end{figure}

Now, we study two different initial conditions: one with a cylindrical shape
\begin{equation}
	\label{eqn:cyliinit}
	u(x, z, 0) = \left\{
		\begin{aligned}
			0.9, \quad & x^2 + z^2 < 0.5, \\
			0, \quad &\mathrm{otherwise},
		\end{aligned}
		\right.  \quad (x,z) \in [-1.5, 1.5] \times [-1.5, 1.5],
\end{equation}
and one with a cubic shape 
\begin{equation}
	\label{eqn:cubeinit}
	u(x, z, 0) = \left\{
		\begin{aligned}
			0.9, \quad & x^2 < 0.5,  z^2 < 0.5, \\
			0, \quad &\mathrm{otherwise},
		\end{aligned}
		\right.  \quad (x,z) \in [-1.5, 1.5] \times [-1.5, 1.5].
\end{equation}
The solutions of the MBLE with initial condition (\ref{eqn:cyliinit}) computed using the uniform mesh and the moving mesh are illustrated in Figs. \ref{fig:tangtao2dmbl}. As one can see from Fig. \ref{fig:tangtao2dmbl}, the MBLE generates a clear plateau at the shock front in the $z$-direction as expected. The plateau heights obtained by central flux are generally higher than those obtained by LLFR flux. The plateau height obtained by the moving mesh ($301^2$ points) with central flux is very close to the TW results, and is even more accurate than the plateau heights getting by the uniform mesh with $1001^2$ points. This indicates about $10$ times saving in the spatial grids, which is especially useful when dealing with 3D computations. 

\begin{figure}[!htbp]
	\begin{center}
		{\includegraphics[width=2.5in] {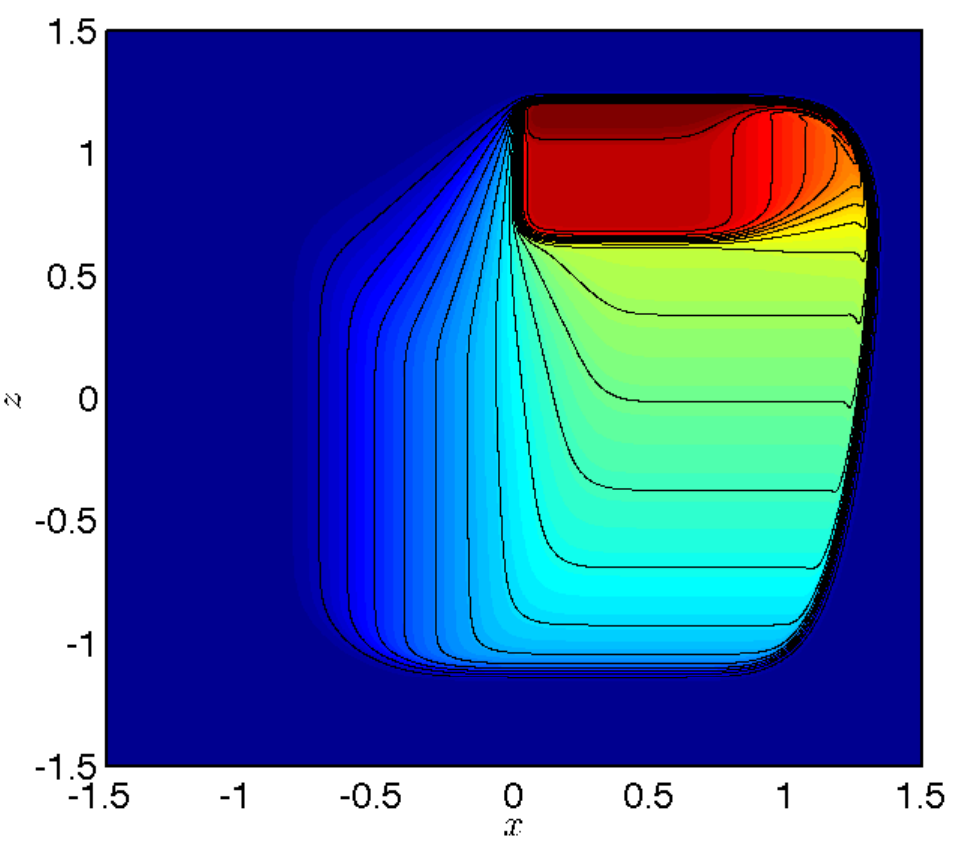}} \quad
		{\includegraphics[width=2.5in] {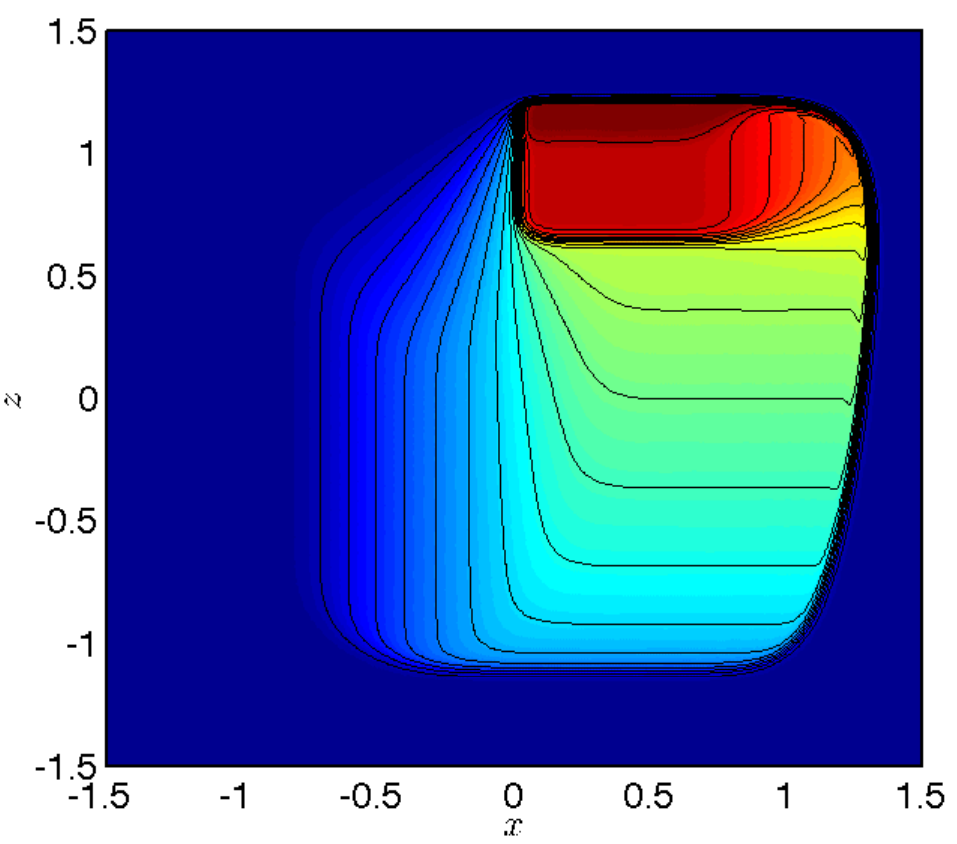}} \\
		\subfigure[Plateau height = $0.9670$]
		{\includegraphics[width=2.5in] {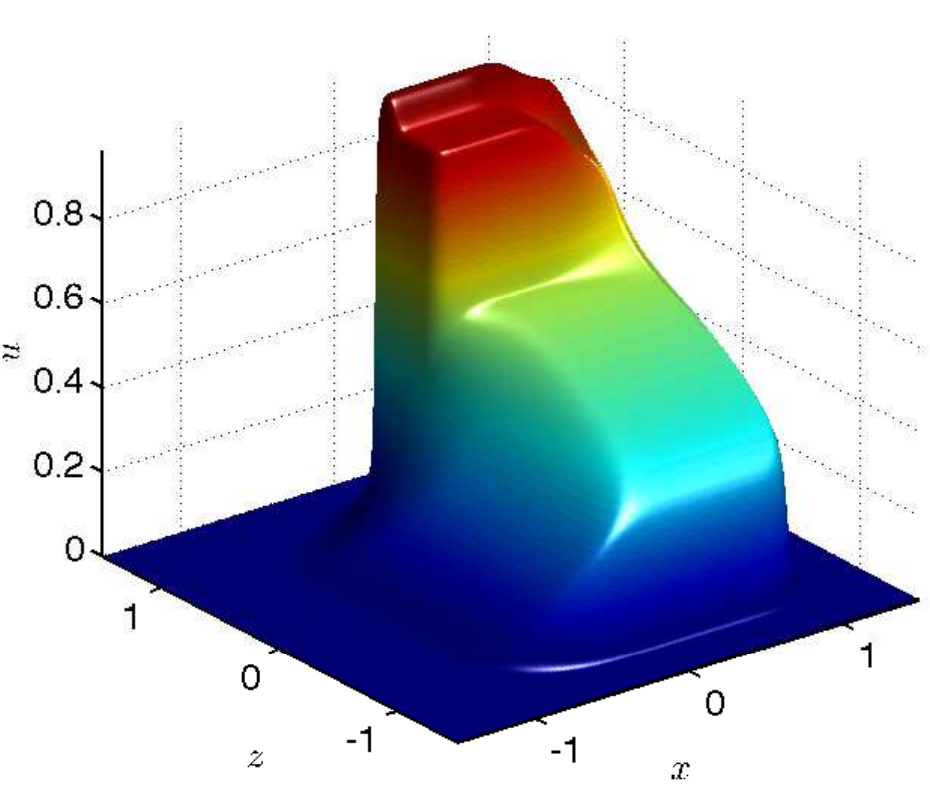}}  \quad 
		\subfigure[Plateau height = $0.9695$]
		{\includegraphics[width=2.5in] {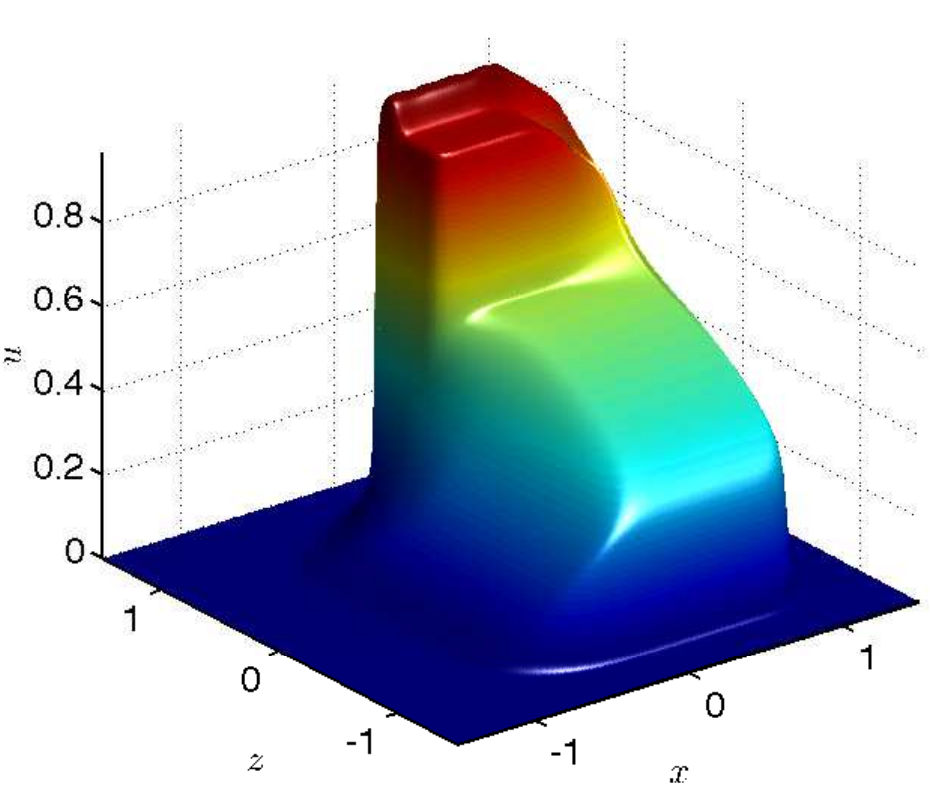}}
		\caption{Example 3-2 with initial condition (\ref{eqn:cubeinit}) at $t = 0.5$: solutions obtained by the uniform mesh and the adaptive mesh. Left column: top and 3D views on uniform mesh ($1001^2$ points) with LLFR flux; right column: top and 3D views on moving mesh ($301^2$ points, $\sigma = 2, \tau_x = \tau_z = 1, \kappa = 0.6$) with central flux. \label{fig:tangtao2dmblrect}}
	\end{center}
\end{figure}

Since the LLFR flux performs better than the central flux in the uniform mesh situation, and the central flux performs better than the LLFR flux in the moving mesh situation, thus in Fig. \ref{fig:tangtao2dmblrect} we show the results with initial condition (\ref{eqn:cubeinit}) for the above choices. Similarly to the previous case of the initial condition (\ref{eqn:cyliinit}), the non-monotone plateaus are located near the shock front in the $z$-direction and become thinner and lower along the positive $x$-direction because of the rarefaction waves created by the flux $F(u)$. Again, the moving mesh method gets a more accurate plateau height than the uniform mesh method. 

\noindent\textbf{Example 4.} In the last example we simulate the finger phenomenon using the RNERE with the Brooks-Corey model. 
Ref. \cite{bauters2000soil} presented snapshots of the finger phenomenon for water infiltrating into $20/30$ sand. In this example, we use the RNERE (\ref{eqn:rnere}) and the Brooks-Corey model to generate a single finger numerically. The physical parameters of the $20/30$ sand \cite{dicarlo2004experimental,schroth1996characterization} as well as the constants and the Brooks-Corey model \cite{brooks1966properties} are listed in Table \ref{tab:2030sand} and Table \ref{tab:model}.
\begin{table}
	\caption{\label{tab:2030sand}Physical parameters for $20/30$ sand.}
	\center
	\begin{tabular}{|cccccc|ccc|}\hline
		{}& {} & {} & \multicolumn{3}{c|}{Drainage}&\multicolumn{3}{c|}{Imbibition}\\\cline{4-9}
		{Sand} & $\kappa~\mathrm{[m~s^{-1}]}$ & $\phi$~[-] & $u_{re}$~[-] & $\lambda$~[-]& $p_d ~\text{[Pa]}$ & $u_{re}$~[-] & $\lambda$~[-] & $p_d ~\text{[Pa]}$\\\hline
		$20/30$ & $2.5\times 10^{-3}$ & $0.35$ & $0$ & $5.57$ & $850$ & $0$ & $5$ & $490$ \\\hline
	\end{tabular}
\end{table}
\begin{table}
	\caption{\label{tab:model}Constants and the Brooks-Corey model.}
	\center
	\begin{tabular}{|c|cc|}\hline
		Density $\mathrm{[kg~m^{-3}]}$ & $\rho_w = 998.21$ & $\rho_n = 1.2754$  \\
		Viscosity $\mathrm{[kg~m^{-1} s^{-1}]}$ & $\mu_w = 1.002\times 10^{-3}$ & $\mu_n = 1.82\times 10^{-5}$  \\
		Mobility $\mathrm{[m~s~kg^{-1}]}$ & $\lambda_w = \frac{K k_{rw}}{\mu_w}$ & $\lambda_n = \frac{K k_{rn}}{\mu_n}$ \\
		Constants & $g = 9.81 ~\mathrm{[m~s^{-2}]}$  & $K = \frac{\kappa \mu_w}{\rho_w g}$ $~\mathrm{[m^2]}$\\\hline
		& Capillary pressure  &{Relative permeability } \\\hline
		& $u_e = \frac{u - u_{re}}{1 - u_{re}}$ &  $k_{rw} = u_e^{\frac{2+ 3 \lambda}{\lambda}}$ \ \\
		\raisebox{1.6ex}[0pt]{Brooks-Corey model} & {
			$p_c = p_d u_e^{-\frac{1}{\lambda}},~~\mathrm{for}~p_c > p_d$}& $k_{rn} = (1-u_e)^2 (1-u_e^{\frac{2+\lambda}{\lambda}})$ \\\hline
	\end{tabular}
\end{table}

\begin{figure}[!htbp]
	\begin{center}
		{\includegraphics[width=2.5in] {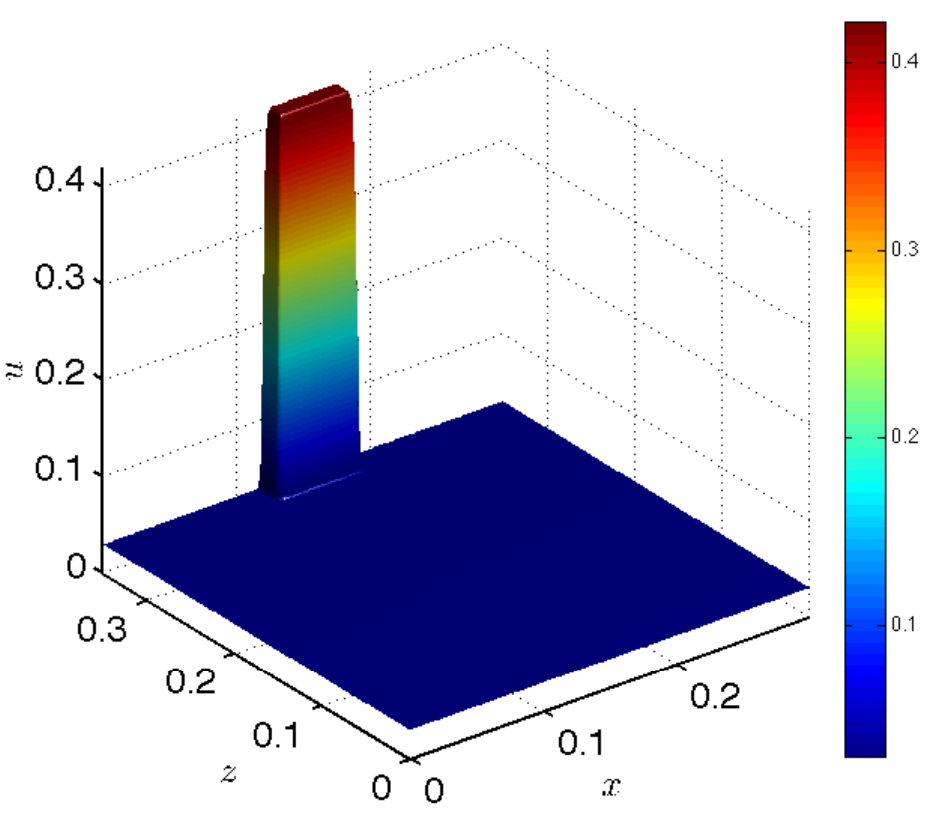}}
		{\includegraphics[width=2.5in] {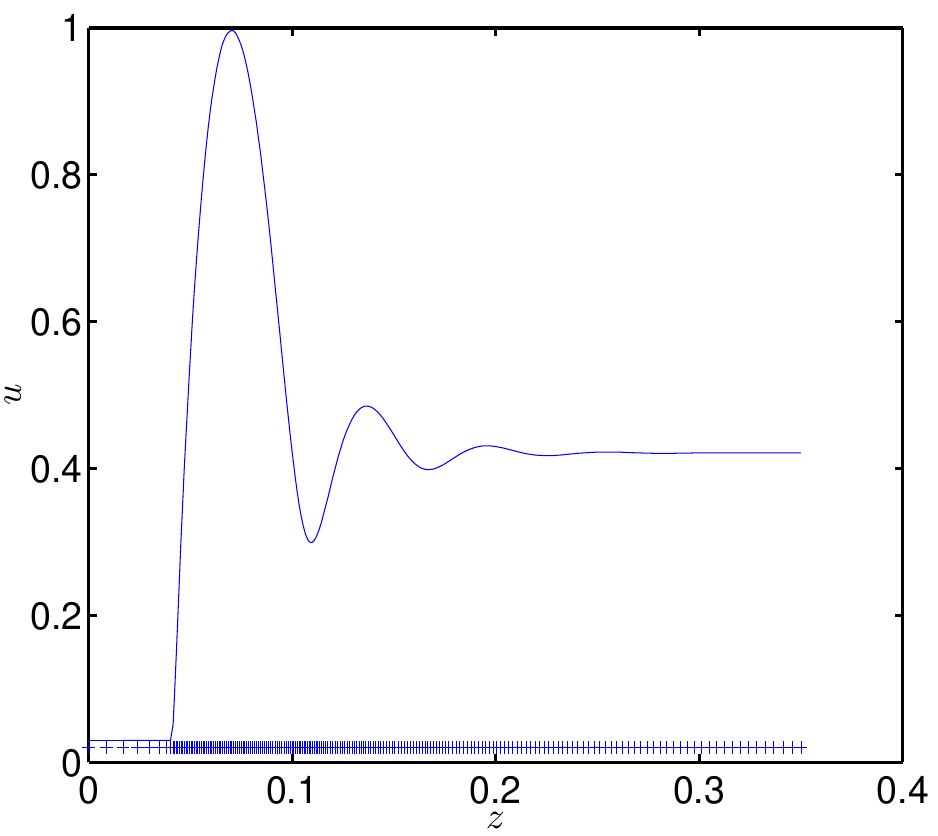}} \quad \\
		{\includegraphics[width=2.5in] {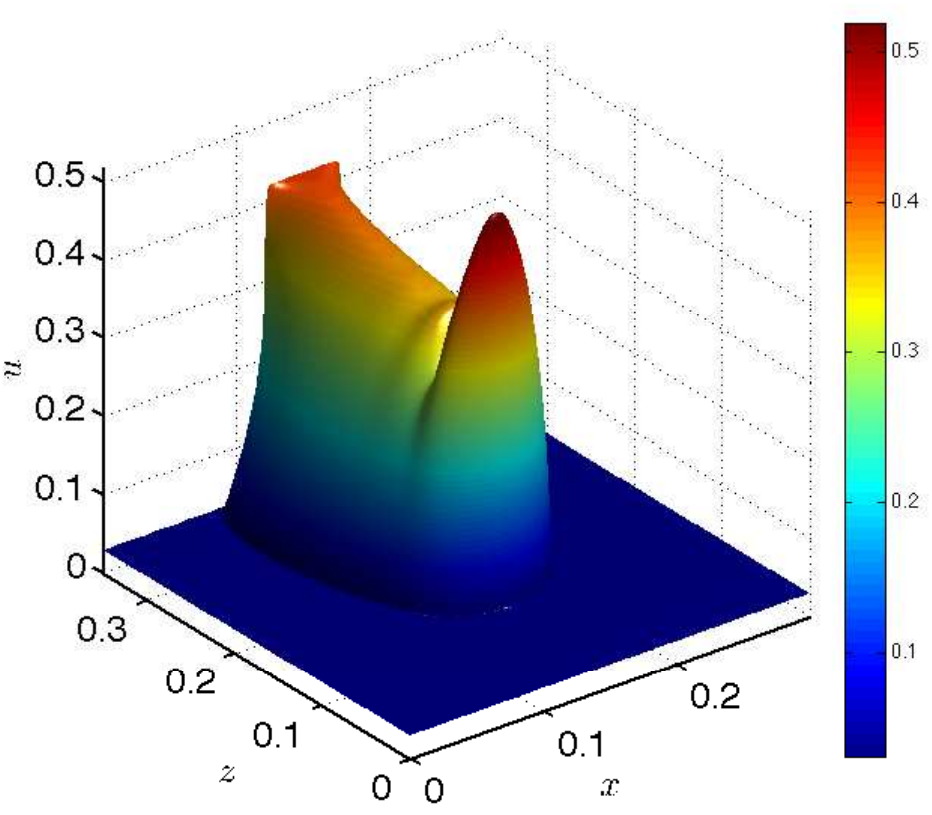}}
		{\includegraphics[width=2.5in] {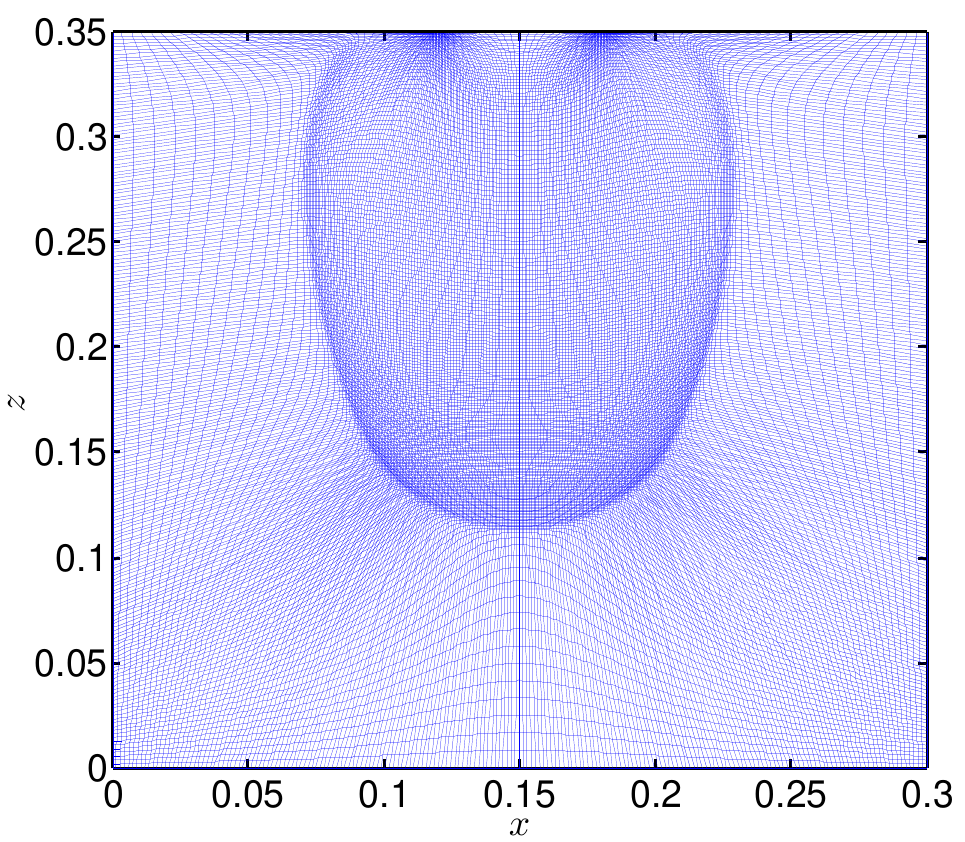}}
		\caption{Example 4 at $t = 350$, top left: 1D moving mesh solution with $201$ points; top right: 2D initial saturation; bottom left: 2D moving mesh solution with $201^2$ points; bottom right: 2D moving mesh with $201^2$ points (right). \label{fig:dicarlore}}
	\end{center}
\end{figure}

Consider the physical domain $[0, 0.3] \times [0, 0.35] [\mathrm{m}]$, let $u_- = 0.03$ and $u_+ = 0.4210$, we take the initial condition as
\begin{equation}
	\begin{aligned}
		%\label{eqn:dicarloreinit}
		u(x, z, 0) =&  u_- + \frac{1}{8}(u_+ - u_-) [ \big(1.0 - \tanh(\frac{200}{x_{max}-x_{min}}(x - 0.18))\big) \\
			& \times \big(1.0 + \tanh(\frac{200}{x_{max}-x_{min}}(x - 0.12))\big)  \\
		& \times \big(1.0 + \tanh(\frac{200}{z_{max}-z_{min}} (z -0.95 (z_{max}-z_{min})))\big)].
	\end{aligned}\nonumber
\end{equation}
The initial saturation is presented is Fig. \ref{fig:dicarlore} (a). In this simulation, we use a Dirichlet boundary condition on the upper boundary $z = 0.35$, a Neumann boundary condition on the lower boundary $z = 0$, and a periodic boundary condition on the vertical boundaries.

Fig. \ref{fig:dicarlore} (c,d) illustrate the results for the numerical simulation at $t = 350$. Along the tail region, in contrast to the 1D simulation in Fig. \ref{fig:dicarlore} (b), the saturation profile for the 2D simulation decreases in the $z$-direction from the upper flow boundary and the overshoot saturation is lower than the value in 1D. This indicates that the lateral flow caused by the pressure gradient greatly influences the saturation profile. In real experiments, as is explained by \cite{glass1989mechanism}, hysteresis is responsible for controlling the finger's sideways growth. In order to simulate realistic fingers, capillary pressure hysteresis has to be incorporated. Since our interest in this work is the dynamic capillary pressure effect, we would like to refer the interested readers to the discussions and simulations considering hysteresis in \cite{nieber2003non,sander2008dynamic,chapwanya2010numerical}.

\section{Conclusions} \label{sec:conclusions}
In the present work, we considered two types of non-equilibrium equations corresponding to the dynamic capillary pressure in porous media. We described the traveling waves for the relaxation non-equilibrium Richards equation and modified Buckley-Leverett equation, and the stability theory of the RNERE was verified by solving the governing equation numerically. Then we introduced a moving mesh finite difference method which is based on the quasi-Lagrangian formulation to approximate the RNERE and MBLE. The numerical scheme was tested on a suite of numerical experiments and showed to be robust. It enabled us to characterize the dynamic capillary effect in some 1D and 2D examples. In particular, we found that the moving mesh method performed much better than the uniform grid method and the central flux with time-dependent curvature monitor is more suitable for the simulating of flows in porous media.

Future work would extend the method of this paper to simulate the finger phenomenon incorporating both dynamic capillary pressure and capillary pressure hysteresis. This will improve the profile of the 2D finger by damping the lateral flow.

\section*{Acknowledgements} \label{sec:ack}
The research of H. Zhang was funded by the China Scholarship Council (No. 201503170430).
\bibliographystyle{elsarticle-num}
\bibliography{reference}

\begin{thebibliography}{10}
\bibitem{eliassi2003porous}
M.~Eliassi, R.~J. Glass, On the porous continuum-scale modeling of
  gravity-driven fingers in unsaturated materials: Numerical solution of a
  hypodiffusive governing equation that incorporates a hold-back-pile-up
  effect, Water resources research 39~(6).
\end{thebibliography}
%%\bibliography{ref}%
%\begin{thebibliography}{99}
%\bibitem{15} A. Latto, H. Resnikoff and E. Tenenbaum, The Evaluation of Connection coefficients
%of compactly supported wavelets, in Proceedings of the French-USA
%Workshop on Wavelets and Turbulence, Springer-Verlag, New York,
%1991.
%
%\end{thebibliography}
\end{document}